\documentclass[traditabstract]{aa}

\usepackage{natbib}
\usepackage{graphicx}
\usepackage[bookmarks=false, colorlinks=true, citecolor=blue, linkcolor=blue]{hyperref}
\DeclareRobustCommand{\ion}[2]{%
\relax\ifmmode
\ifx\testbx\f@series
{\mathbf{#1\,\mathsc{#2}}}\else
{\mathrm{#1\,\mathsc{#2}}}\fi
\else\textup{#1\,{\mdseries\textsc{#2}}}%
\fi}

\newcommand{\Hii}{\ion{H}{ii}}
\newcommand{\Spitzer}{\textit{Spitzer}}
\newcommand{\Herschel}{\textit{Herschel}}
\def\micron{\hbox{\,$\mu$m}}
\newcommand{\Lsun}{\hbox{$L_{\rm \odot}$}}
\newcommand{\Msun}{\hbox{$M_{\rm \odot}$}}

\newcommand{\degree}{\ensuremath{^\circ}}
\newcommand\nodata{ ~$\cdots$~ }

\newsavebox\boxmap

\bibpunct{(}{)}{;}{a}{}{,}

\titlerunning{Molecular outflow of ESO320-G030}
\authorrunning{Pereira-Santaella et al.}

\begin{document}

\title{High-velocity extended molecular outflow in the star-formation dominated luminous infrared galaxy ESO~320-G030}
\author{M. Pereira-Santaella\inst{\ref{inst1}, \ref{inst2}, \ref{inst3}} \and L. Colina\inst{\ref{inst1}, \ref{inst2}} \and S. Garc\'ia-Burillo\inst{\ref{inst4}} \and A. Alonso-Herrero\inst{\ref{inst1}}  \and S. Arribas\inst{\ref{inst1}, \ref{inst2}} \and S. Cazzoli\inst{\ref{inst1}, \ref{inst2}} \and  B. Emonts\inst{\ref{inst1}, \ref{inst2}} \and J. Piqueras L\'opez\inst{\ref{inst1}, \ref{inst2}} \and P. Planesas\inst{\ref{inst4}}  \and T. Storchi Bergmann\inst{\ref{inst5}} \and A. Usero\inst{\ref{inst4}}  \and M. Villar-Mart\'in\inst{\ref{inst1}, \ref{inst2}}
}
\institute{Centro de Astrobiolog\'ia (CSIC/INTA), Ctra de Torrej\'on a Ajalvir, km 4, 28850, Torrej\'on de Ardoz, Madrid, Spain \label{inst1} \and
ASTRO-UAM, UAM, Unidad Asociada CSIC\label{inst2} 
\and
Department of Physics, University of Oxford, Keble Road, Oxford OX1 3RH, UK \label{inst3} \\ \email{miguel.pereira@physics.ox.ac.uk} 
\and
Observatorio Astron\'omico Nacional (OAN-IGN)-Observatorio de Madrid, Alfonso XII, 3, 28014, Madrid, Spain\label{inst4}
\and
Universidade Federal do Rio Grande do Sul, Instituto de F\'isica, CP 15051, Porto Alegre 91501-970, RS, Brazil\label{inst5}  
}

\abstract{
We analyze new high spatial resolution ($\sim$60\,pc) ALMA CO(2--1) observations of the isolated luminous infrared galaxy ESO~320-G030 ($d=48$\,Mpc) in combination with ancillary \textit{HST} optical and near-IR imaging as well as VLT\slash SINFONI near-IR integral field spectroscopy. 
We detect a high-velocity ($\sim$450\,km\,s$^{-1}$) spatially resolved (size$\sim$2.5\,kpc; dynamical time $\sim$3\,Myr) massive ($\sim$10$^7$\,\Msun; $\dot{M}\sim$2--8\,\Msun\,yr$^{-1}$) molecular outflow originated in the central $\sim$250\,pc.
We observe a clumpy structure in the outflowing cold molecular gas with clump sizes between 60 and 150\,pc and masses between 10$^{5.5}$ and 10$^{6.4}$\,\Msun. 
The mass of the clumps decreases with increasing distance, while  the velocity is approximately constant. Therefore, both the momentum and kinetic energy of the clumps decrease outwards. 
In the innermost ($\sim$100\,pc) part of the outflow, we measure a hot-to-cold molecular gas ratio of 7$\times$10$^{-5}$, which is similar to that measured in other resolved molecular outflows. We do not find evidence of an ionized phase in this outflow.
The nuclear IR and radio properties are compatible with strong and highly obscured star-formation ($A_{\rm k}\sim4.6$\,mag; ${\rm SFR}\sim15$\,\Msun\,yr$^{-1}$). 
We do not find any evidence for the presence of an active galactic nucleus.
We estimate that supernova explosions in the nuclear starburst ($\nu_{\rm SN}\sim0.2$\,yr$^{-1}$) can power the observed molecular outflow. The kinetic energy and radial momentum of the cold molecular phase of the outflow correspond to about 2\% and 20\%, respectively, of the supernovae output.
The cold molecular outflow velocity is lower than the escape velocity, so the gas will likely return to the galaxy disk. The mass loading factor is $\sim$0.1--0.5, so the negative feedback due to this star-formation powered molecular outflow is probably limited.
}


\maketitle

\section{Introduction}\label{s:intro}

Theoretical models predict that massive gas outflows, driven by starbursts or active galactic nuclei (AGN), are fundamental actors in shaping the observed properties of galaxies (e.g., galaxy mass function, mass-metallicity relation).
This is because massive outflows can regulate both the accretion rate of the central supermassive black hole and the star-formation (SF) activity, but also because they can redistribute dust and metals over kpc scales (e.g., \citealt{Veilleux2005, Narayanan2008, Hopkins2012}).

Gas outflows have a multiphase (ionized, neutral atomic, and molecular) structure \citep{Veilleux2005, Hopkins2012}. The ionized and neutral atomic phases have been studied over the past 25 years using optical and ultraviolet spectral features (e.g., \citealt{Heckman1990, Veilleux1995, Rupke2008, Arribas2014, Heckman2015, Cazzoli2016}). 
Just recently, thanks to improved capabilities of millimeter observatories, many works have focused on the molecular phase of the outflow, which dominates the outflowing gas mass, energy, and momentum in most cases (e.g., \citealt{Feruglio2010, Tsai2012, Bolatto2013ngc253, Cicone2014, GarciaBurillo2014, GarciaBurillo2015}). These studies show that massive molecular outflows are ubiquitous in AGN and starbursts and that they might have an important impact in the evolution of their host galaxies.

Multi-transition and multi-specie studies of the molecular phase of the outflow shed some light on the poorly known physical and chemical properties of this phase. For instance, the isotopic O abundance measured using the far-IR OH absorption in the outflow of Mrk~231 suggests that the outflowing gas has been processed by advanced starbursts \citep{Fischer2010}. In addition, the HCN, HNC, and HCO$^+$ emissions indicate that the molecular gas in that outflow can be chemically differentiated while being compressed and fragmented by shocks \citep{Aalto2012, Lindberg2016}. Actually, the analysis of cold and hot molecular gas tracers (e.g., CO and near-infrared H$_2$ transitions) in various objects indicates that the outflowing molecular gas is continuously heated by shocks \citep{Dasyra2014, Emonts2014}. 

{Similarly, spatially resolved observations of outflows at sub-kpc scales (e.g., \citealt{Bolatto2013ngc253, Emonts2014, Sakamoto2014, GarciaBurillo2014, Salak2016}) are essential to study the outflowing molecular gas. 
The clumpy distribution of the molecular gas in the outflow, as shown in this work, and the physical properties of these clumps provide a benchmark for models and simulations to establish how the molecular gas evolves in the outflow (e.g., \citealt{Nath2009, Zubovas2014b}) and also to better constrain the global effect of the outflow in the host galaxy.
}

Many of these {outflow} studies analyze local luminous and ultraluminous infrared galaxies (LIRGs and ULIRGs respectively) because they are the most extreme examples of starbursts (and AGN) in the local Universe. LIRGs have infrared (IR) luminosities $>$10$^{11}$\,\Lsun, while ULIRGs have $L_{\rm IR}>10^{12}$\,\Lsun. These IR luminosities, if produced solely by star-formation, are equivalent to star-formation rates (SFR) of $>$15 and $>$150\,\Msun\,yr$^{-1}$ , respectively {(see Table 1 of \citealt{Kennicutt2012}).}
These local U\slash LIRGs are considered as local counterparts of high-$z$ starburst with similar or higher IR luminosities (e.g., \citealt{Muzzin2010}).
Therefore, they are perfect targets for detailed studies at high spatial resolution of the processes, for example outflows, taking place in their high-$z$ counterparts.

In this paper, we analyze new Atacama Large Millimeter/submillimeter Array (ALMA) CO(2--1) high spatial resolution ($\sim$60\,pc) observations of the local ($d=48$\,Mpc; {scale of }240\,pc\,arcsec$^{-1}$) LIRG ESO~320-G030 ($\log L_{\rm IR}\slash \Lsun = 11.3$; also known as IRAS~F11506-3851). We combine the CO(2--1) data (cold molecular gas tracer) with available near-IR integral field spectroscopy (IFS) to trace the hot molecular gas as well as the ionized gas \citep{Piqueras2012}.
This galaxy is an isolated spiral galaxy with an ordered velocity field \citep{Bellocchi2013} hosting a strong starburst \citep{AAH06s}.
Its nuclear activity is classified as \ion{H}{ii} from optical spectroscopy \citep{vandenBroek1991, Pereira2011} and there is no evidence of an AGN in this galaxy based on its mid-IR and X-ray emissions \citep{Pereira2010c, Pereira2011}.
It also host an OH megamaser \citep{Norris1986}, which are associated with compact starbursts most of the times \citep{Baan2006, Zhang2014}.
In addition, a massive outflow of neutral atomic gas is already detected in this object using optical IFS \citep{Cazzoli2014, Cazzoli2016}.

This paper is organized as follows: we describe the observations and data reduction in Section \ref{s:data}. The analysis of the gas and stellar morphology and kinematics are presented in Sections \ref{s:morphology} and \ref{s:vfield}. In Section \ref{s:outflow}, we describe the structure and physical properties of the resolved molecular outflow. In Section \ref{s:nature}, we investigate the nature of the compact central source of this galaxy. The main results are discussed in Section \ref{s:discussion} and, finally, in Section \ref{s:conclusions}, we summarize the main findings of the paper.

\section{Observations and data reduction}\label{s:data}

\begin{figure*}
\centering
\includegraphics[width=0.9\textwidth]{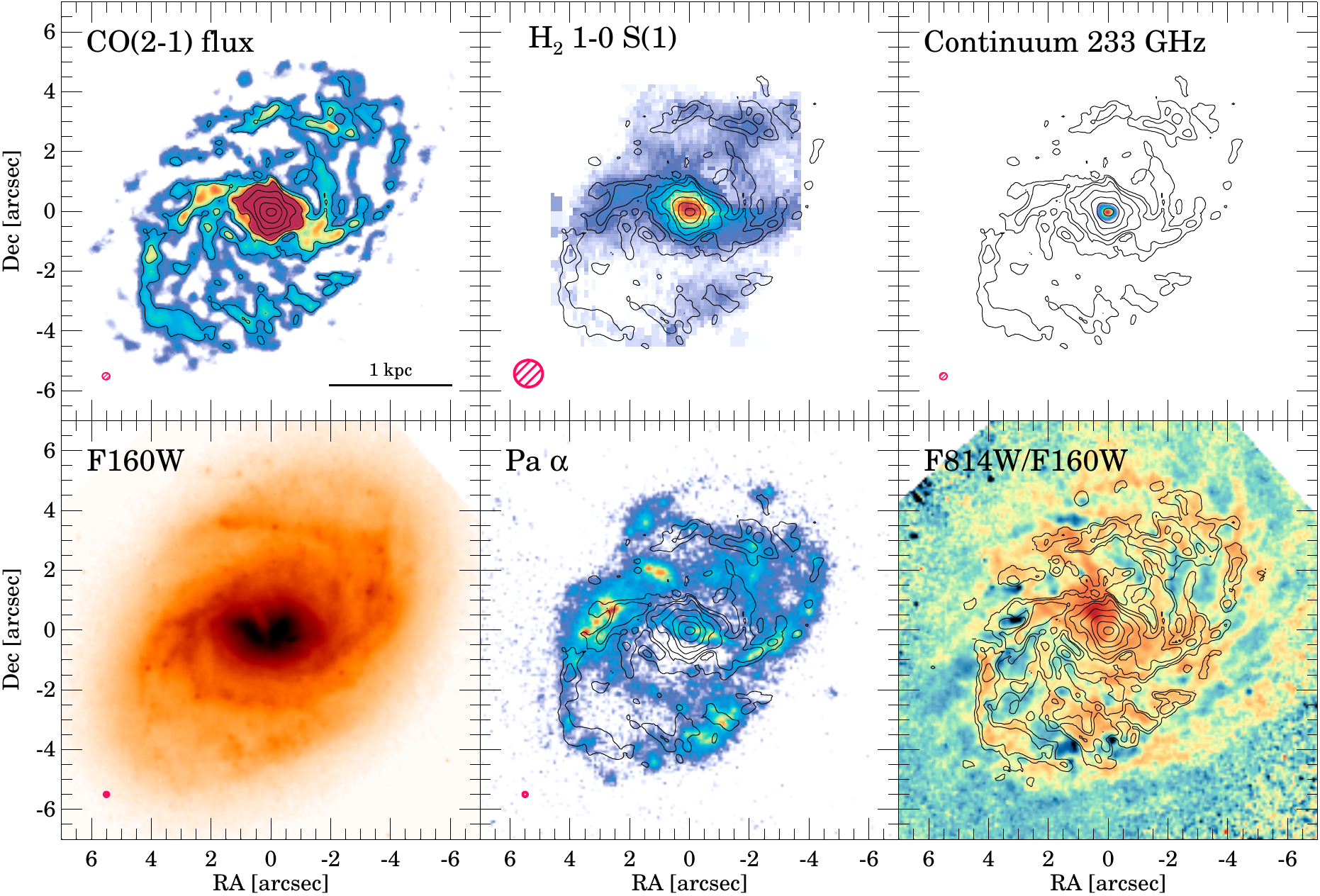}
\caption{\small The top row shows the ALMA $^{12}$CO(2--1), VLT\slash SINFONI H$_2$ 1--0 S(1) 2.12\micron, and ALMA 233\,GHz continuum (rest frequency) maps of ESO\,320-G030. The contours in all the images correspond to the CO(2--1) emission in logarithmic steps (1, 2, 4, 8, 16, 32, 64)$\times$0.43\,Jy\,km\,s$^{-1}$\,beam$^{-1}$.
The lower panels show the \textit{HST}\slash NICMOS F160W and continuum subtracted Pa$\alpha$ maps and the \textit{HST} F814W to F160W ratio to highlight obscured regions (red colors indicate more obscured regions). In the last panel we added an extra contour level at 0.5$\times$0.43\,Jy\,km\,s$^{-1}$\,beam$^{-1}$.
{The red hatched ellipses indicate the beam size of each image. For the ALMA CO(2--1) and the 233\,GHz continuum images, the beam size is 0\farcs25$\times$0\farcs23, PA 89\degree. For the SINFONI H$_2$ 1--0 S(1) image, the beam FWHM is 0\farcs6, and for the \textit{HST}\slash NICMOS data, the beam is $\sim$0\farcs15.}
\label{fig_maps}}
\end{figure*}

\subsection{$^{12}$CO(2--1) ALMA data}\label{s:co_alma}

Band 6 ALMA observations of ESO~320-G030 were obtained on 8 December 2014 and 18 July 2015 using compact and extended array configurations with 35 and 39 antennas, respectively, as part of the project 2013.1.00271.S (PI: L. Colina). 
On-source integration times were 8 and 17\,min, respectively. Both observations were single pointing centered at the nucleus of ESO~320-G030. The extended configuration had baselines between 15.1\,m and 1.6\,km while for the compact configuration the baselines ranged between 15.0\,m and 348\,m. For these configurations the maximum recoverable scales are {$\sim$10\arcsec}.

Two spectral windows of 1.875\,GHz bandwidth (0.49\,MHz~$\sim$ 0.6\,km\slash s channels) were centered at the sky frequencies of $^{12}$CO(2--1) (228.2\,GHz) and CS(5--4) (242.4\,GHz). In addition, two continuum spectral windows were set at 230.4 and 245.2\,GHz ($\sim$1.3\,mm). We excluded the CS(5--4) transition from our analysis because its redshifted frequency coincides with a variable atmospheric feature.

The two datasets were calibrated using the standard ALMA reduction software CASA (v4.2.2; \citealt{McMullin2007}). For the amplitude calibration we used J1147-3812, assuming a flux density of 0.817\,Jy at 236.5\,GHz, and Ganymede, using the Butler-JPL-Horizons 2012 model, for the extended and compact configurations, respectively.
The $uv$ visibilities of each observation were converted to a common frequency reference frame (kinematic local standard of rest; LSRK) and then combined. We checked that the amplitudes of the baselines in common for both array configurations were in good agreement.
For the CO(2--1) data, the continuum was fitted with a constant using the line free channels and subtracted in the $uv$ plane. In the final data cubes, we used 4\,MHz channels ($\sim$5\,km\,s$^{-1}$) {to increase the signal-to-noise ratio} and 256$\times$256 pixels of 0\farcs06. For the cleaning of both the CO(2--1) and continuum data, we used the Briggs weighting with a robustness parameter of 0.5 \citep{Briggs1995PhDT} which provided a beam with a full-width half-maximum (FWHM) of 0\farcs25$\times$0\farcs23 ($\sim$60\,pc$\times$55\,pc) with a position angle (PA) of 89\degree.
A mask derived from the observed CO(2--1) emission in each channel was used during the clean process. The achieved 1$\sigma$ sensitivity is $\sim$1\,mJy\,beam$^{-1}$ in the CO(2--1) cube for the 4\,MHz channels and $\sim$100\,$\mu$Jy\,beam$^{-1}$ in the continuum images. We applied the primary beam (FWHM=27\arcsec) correction to the data.

The integrated CO(2--1) flux in the {analyzed ALMA field of view} (15\arcsec$\times$15\arcsec) is 640\,Jy\,km\,s$^{-1}$ with a flux calibration uncertainty about 15\%. To our knowledge, no single dish CO(2--1) observations of this galaxy exist. A CO(1--0) flux of 180\,Jy\,km\,s$^{-1}$ was reported by \citet{Mirabel1990} using the 15\,m SEST telescope ({44}\arcsec\ beam size). Assuming an $r_{21}$ ratio of 0.9 \citep{Bolatto2013}, the expected CO(2--1) flux would be $\sim$650\,Jy\,km\,s$^{-1}$, which is slightly higher than our measured flux. Although the single dish data covers a larger area and the $r_{21}$ ratio is somewhat uncertain, this similarity suggests that we recover the majority of the CO(2--1) flux combining both the compact and the extended ALMA array configurations.

\subsection{Ancillary data}

\subsubsection{Optical and near-IR \textit{HST} imaging}

We used the continuum subtracted narrow-band Pa$\alpha$ and the broad-band F160W ($\lambda_{\rm c}=$1.60\micron, FWHM=0.34\micron) images obtained with \textit{HST}\slash NICMOS presented by \citet{AAH06s} to trace the SFR and the distribution of the stellar mass, respectively. 
From the Mikulski Archive for Space Telescopes (MAST), we downloaded the \textit{HST}\slash ACS reduced images obtained with the filter F814W ($\lambda_{\rm c}=$8012\,\AA, FWHM=1539\,\AA) and the ramp filter FR656N centered at 6634\,\AA\ (redshifted wavelength of H$\alpha$ for this object) with a bandwidth of 46\AA. We used these images to map the global morphology of ESO~320-G030.

\subsubsection{Near-IR \textit{VLT}\slash SINFONI IFS}

We observed this galaxy with the near-IR VLT\slash SINFONI integral field spectrograph in the $J$ and $K$ bands. These observations have been analyzed by \citet{Piqueras2012, Piqueras2016} and  \citet{Cazzoli2014}. They cover a field of view of 8\arcsec$\times$8\arcsec\ centered at the nucleus of the galaxy with a seeing limited angular resolution of $\sim$0\farcs6. 
From these observations, we use the Br$\gamma$ to Br$\delta$ ratio maps to derive the extinction, the H$_2$ 1--0 S(1) 2.12\micron\ transition to trace the presence of hot molecular gas, and the [\ion{Fe}{ii}]1.64\micron\ as a supernovae tracer.

\subsubsection{Far-IR \textit{Herschel} imaging}
Far-IR \Herschel\ imaging of ESO~320-G030 was available in the \Herschel\ archive. We downloaded the PACS \citep{Poglitsch2010PACS} 70, 100, and 160\micron\ images as well as the SPIRE \citep{Griffin2010SPIRE} 250, 350, and 500\micron\ images. We processed the data using the standard \Herschel\ pipeline environment software (HIPE) and created the maps using Scanamorphos \citep{Roussel2012}. The angular resolution of the images varies from 6\arcsec\ to 35\arcsec\ depending on the wavelength. More details on the data reduction can be found in \citet{Pereira2015not}.

\subsubsection{Image alignment}

Figure \ref{fig_maps} shows the ALMA, \textit{HST}\slash NICMOS, and SINFONI maps of ESO~320-G030. 
Since the absolute positioning, i.e. astrometry, of the \textit{HST} images and SINFONI cubes is in general not known to better than $\sim$1\arcsec, the alignment of the different datasets required several steps.
First, we took the ALMA position as reference. The CO(2--1) spiral arms closely resemble the dust lanes observed in the optical F814W and FR656N images, so we used these features to align these two optical images with the ALMA maps. Next, we used the stars in the field of view of both the F814W and F160W images to align the NICMOS images (F160W and Pa$\alpha$). Finally, the SINFONI maps (H$_2$ 1--0 S(1) and Br$\gamma$) were aligned by matching the Br$\gamma$ map (see \citealt{Piqueras2012}) with the NICMOS Pa$\alpha$ image. The uncertainty of the alignment is around 0\farcs15 ($\sim$pixel size of SINFONI).

\section{Global morphology}\label{s:morphology}

\begin{figure}
\centering
\includegraphics[width=0.5\textwidth]{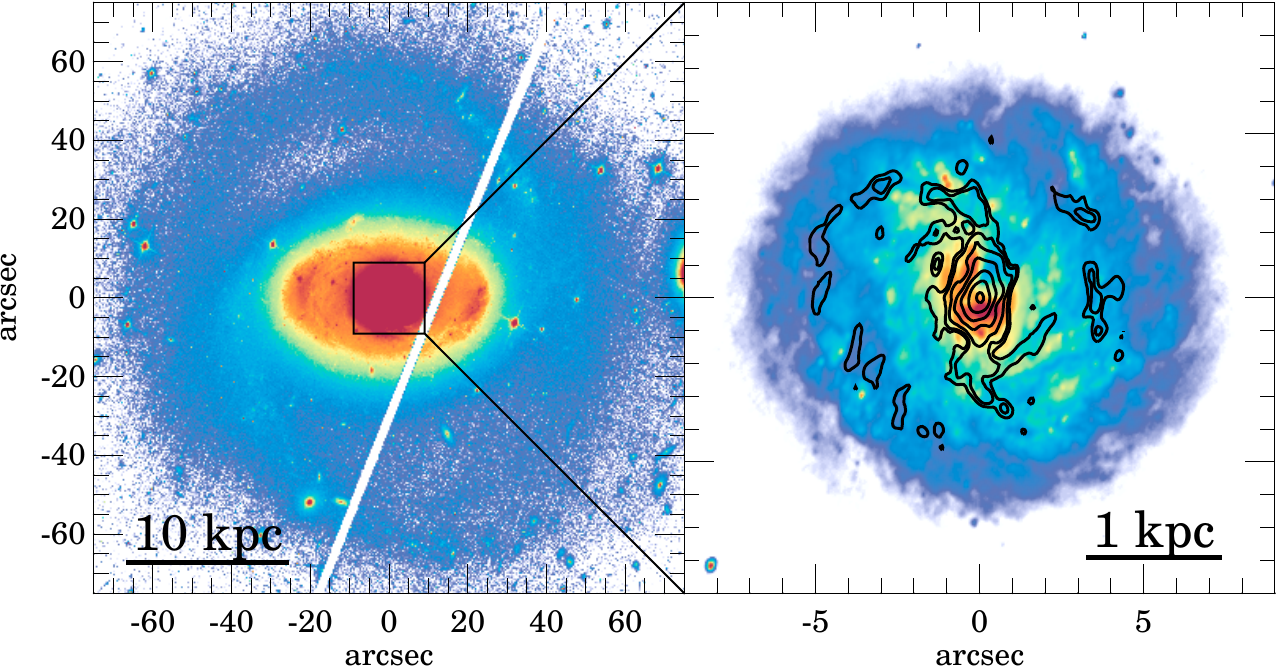}
\caption{\small Deprojected F814W image of ESO~320-G030 assuming an inclination of 43\degree\ and a PA of 133\degree. In the left panel, the central region is saturated to make visible the primary bar and the external spiral arms (pseudoring). The right panel shows the secondary bar and spiral structure present in the central 5\,kpc. The contours represent the deprojected CO(2--1) emission. The diagonal white line in the left panel corresponds to the gap between the two \textit{HST}\slash ACS detectors.
\label{fig_deproj}}
\end{figure}

ESO~320-G030 is an isolated barred spiral galaxy \citep{Buta1986, Greusard2000}.
In the left panel of Figure \ref{fig_deproj}, we show the deprojected\footnote{For the deprojection, we used the PA derived from the kinematic axes (see next Section) and the inclination from the elliptical isophote fitting to the internal spiral structure.} optical image where a large-scale primary bar (semi-major axis $a\sim9$\,kpc) and the outer pseudoring are clearly visible.
A secondary stellar bar ($a\sim1.2$\,kpc) within the primary bar was observed in the near-IR with an internal spiral structure starting from it (\citealt{Greusard2000}; see F160W panel in Figure \ref{fig_maps}). \citet{Greusard2000} measured a PA difference between the primary and secondary stellar bars of 35\degree\ using near-IR $K$-band images. However, the high-angular resolution \textit{HST} and ALMA data (right panel of Figure \ref{fig_deproj}) show that this secondary stellar bar and the elongated molecular structure associated with it are approximately perpendicular to the primary bar. This difference in the measured PA could be due to the lower angular resolution of their near-IR images (seeing$>$1\farcs2) and the effect of the strong dust lanes on those images. 

Almost 60\%\ of the CO(2--1) emission in the field of view is concentrated within the central 500\,pc (see Figure \ref{fig_maps}). This accumulation of gas in the nuclear region is likely driven by the inward gravity torques due to the decoupled secondary bar, since most of it is inside corotation. This is the common scenario reproduced by n-body numerical simulations \citep{Hunt2008}, including gas dissipation, which predict the secondary bar decoupling and the gas inflow to the inner Lindblad resonance of the secondary bar.

The hot molecular gas traced by the H$_2$ 1--0 S(1) emission follows remarkably well the the morphology of CO(2--1) emission. The observed H$_2$ 1--0 S(1) emission is less concentrated in the nucleus than the CO(2--1) emission ($\sim$25\%\ of the integrated emission is produced in the central 500\,pc vs $\sim$60\% for the CO), but this can be due to the large extinction of the nuclear region (see Section \ref{ss:nucsfr}).
On the contrary, the SF traced by Pa$\alpha$ is dominated by the emission from the internal spiral arms. We also note that most of the \Hii\ regions do not exactly coincide with the CO(2--1) emitting regions. 
This agrees with our previous findings in another LIRG, IC~4687 \citep{Pereira2016}, where ALMA observations suggest that the CO and Pa$\alpha$ emitting regions do not generally coincide at 100--200\,pc scales, and therefore the SF law breaks down on these sub-kpc scales. {Similar results are obtained for spiral (e.g., \citealt{Schruba2010}) and interacting galaxies (e.g., \citealt{Zaragoza-Cardiel2014}) as well as for irregular galaxies (e.g., \citealt{Kawamura2009}).}

We used the ratio between the F814W and the F160W images to obtain a qualitative measurement of the extinction. The bottom right panel of Figure \ref{fig_maps} shows the good agreement of this ratio with the CO(2--1) emission.

Finally, the 233\,GHz continuum comes from a slightly resolved source (0\farcs34$\times$0\farcs30) located at the nucleus of this object. We discuss the properties of the nucleus in detail in Section \ref{s:nature}.

\section{Global kinematics}\label{s:vfield}

\begin{figure}
\centering
\includegraphics[width=0.43\textwidth]{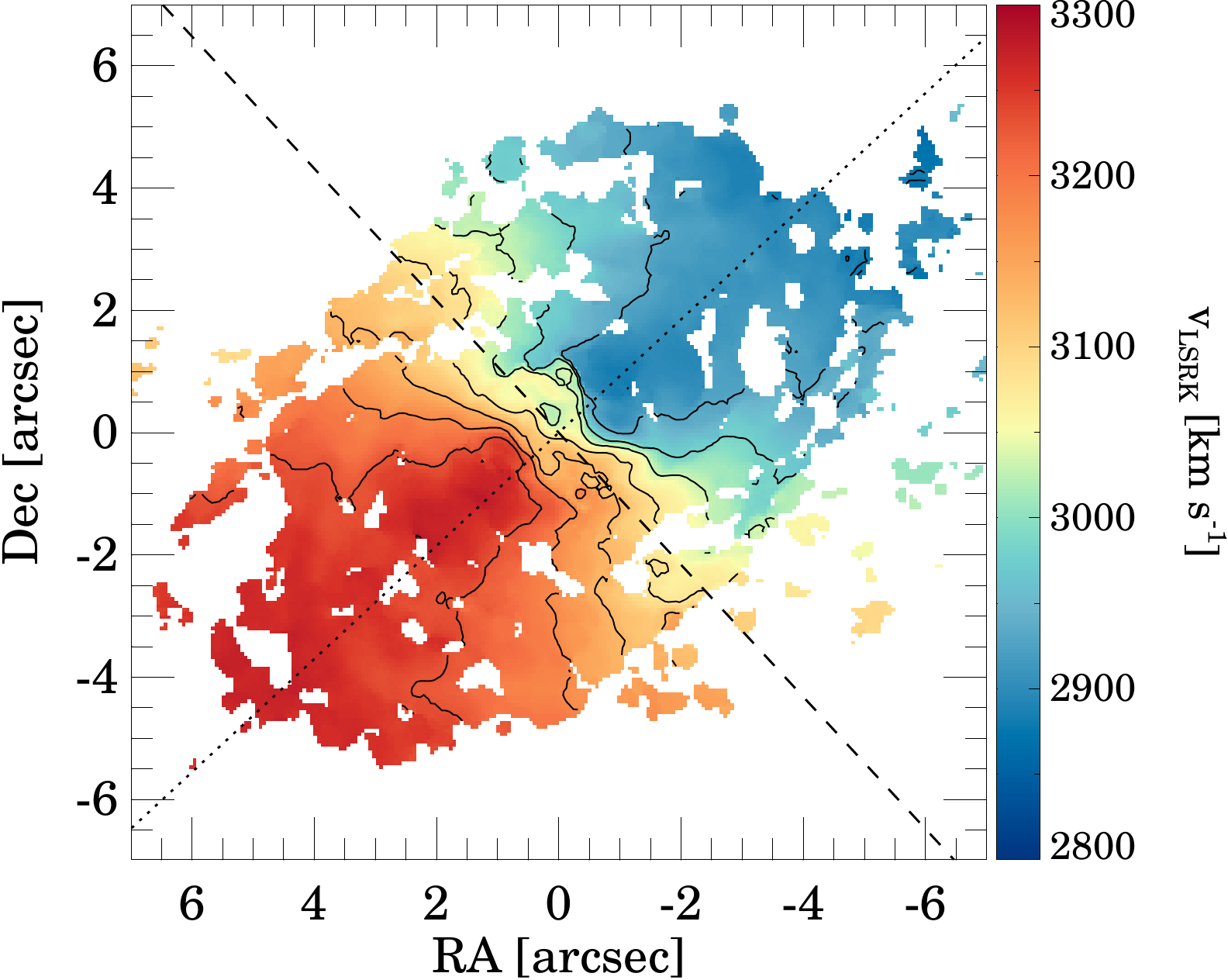}
\caption{\small CO(2--1) isovelocity contours in steps of 50\,km\,s$^{-1}$ from $v$(\textsc{lsrk}) = 2800 to 3300\,km\,s$^{-1}$. The dotted and the dashed lines indicate the major and minor kinematic axes.
\label{fig_isovel}}
\end{figure}

\begin{figure*}
\centering
\includegraphics[width=0.94\textwidth]{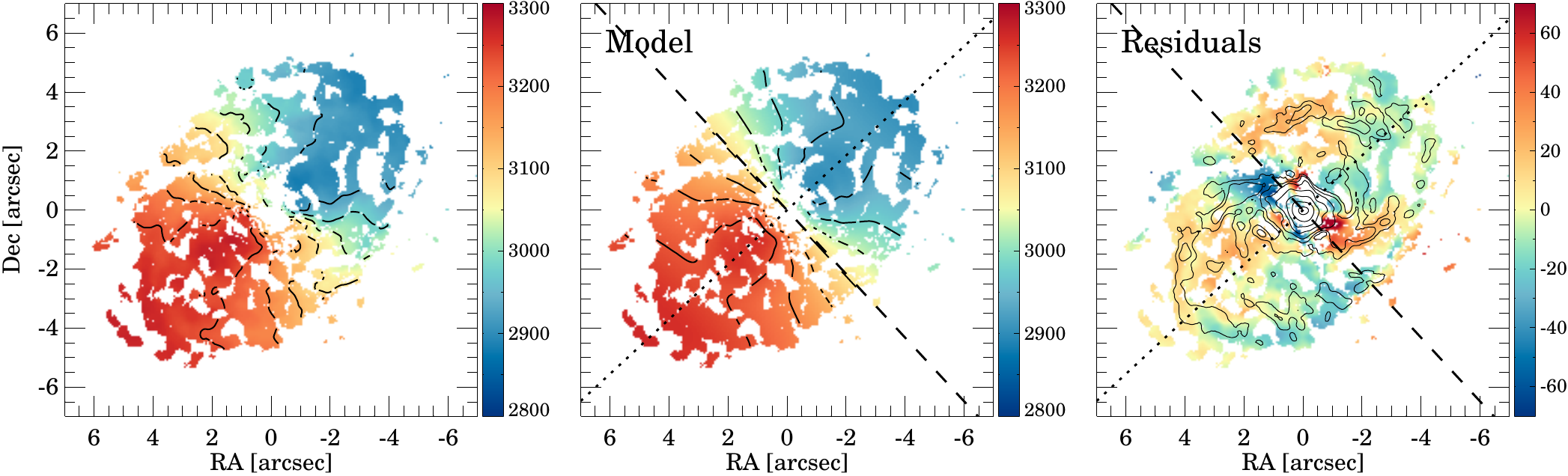}
\caption{\small The left panel shows the CO(2--1) velocity field after excluding those pixels with complex line profiles. The middle panel show the best fitting model (see Section \ref{s:vfield}). In these two panels, the contour levels are set in steps of 50\,km\,s$^{-1}$ from $v$(\textsc{lsrk}) = 2800 to 3300\,km\,s$^{-1}$. The right panel shows the residuals (data -- model) and the CO(2--1) emission contours as in Figure \ref{fig_maps}.
The dotted and the dashed lines indicate the major and minor kinematic axes.
\label{fig_velmodel}}
\end{figure*}

The H$\alpha$ velocity field of this object ($\sim$1\arcsec\ resolution) was studied by \citet{Bellocchi2013}. It shows a very regular pattern with a centrally peaked velocity dispersion which these authors classify as a pure rotating disk. {This regular pattern is also observed using the near-IR Br$\gamma$ and H$_2$ 1--0 S(1) emission lines \citep{Piqueras2012}.}

In Figure \ref{fig_isovel}, we show the velocity field obtained from the first moment of the CO(2--1) emission. It also presents a regular rotating disk pattern, although some deviations are present (see also \citealt{Bellocchi2013} and \citealt{Cazzoli2014}). To quantify them, we modeled the velocity field using a simple kinematic model (e.g., \citealt{Mihalas1981}):
\begin{equation}
v(r,\theta) = \Omega(r)\times \cos(\theta - \theta_0)\times \sin i + v_{\rm sys}
\end{equation}
where $r$ and $\theta$ are the galactocentric distance and azimuthal angle, respectively, $\Omega(r)$ is the azimuthal velocity, $\theta_0$ is the PA of the major kinematic axis, $i$ the inclination angle of the galaxy, and $v_{\rm sys}$ the systemic velocity. 

In the central region, the CO(2--1) emission has complex profiles (see Section \ref{s:outflow}) indicating that the gas kinematic there will not be well reproduced by our simple model. To exclude these regions, we fitted the CO(2--1) profiles pixel by pixel using a Gaussian and rejected all those pixels where the quality of the fit was poor ({reduced} $\chi^2>1.4$).
The left panel of Figure \ref{fig_velmodel} shows the velocity field used for the kinematic model fit with the central pixels excluded.

\begin{figure}
\centering
\includegraphics[width=0.43\textwidth]{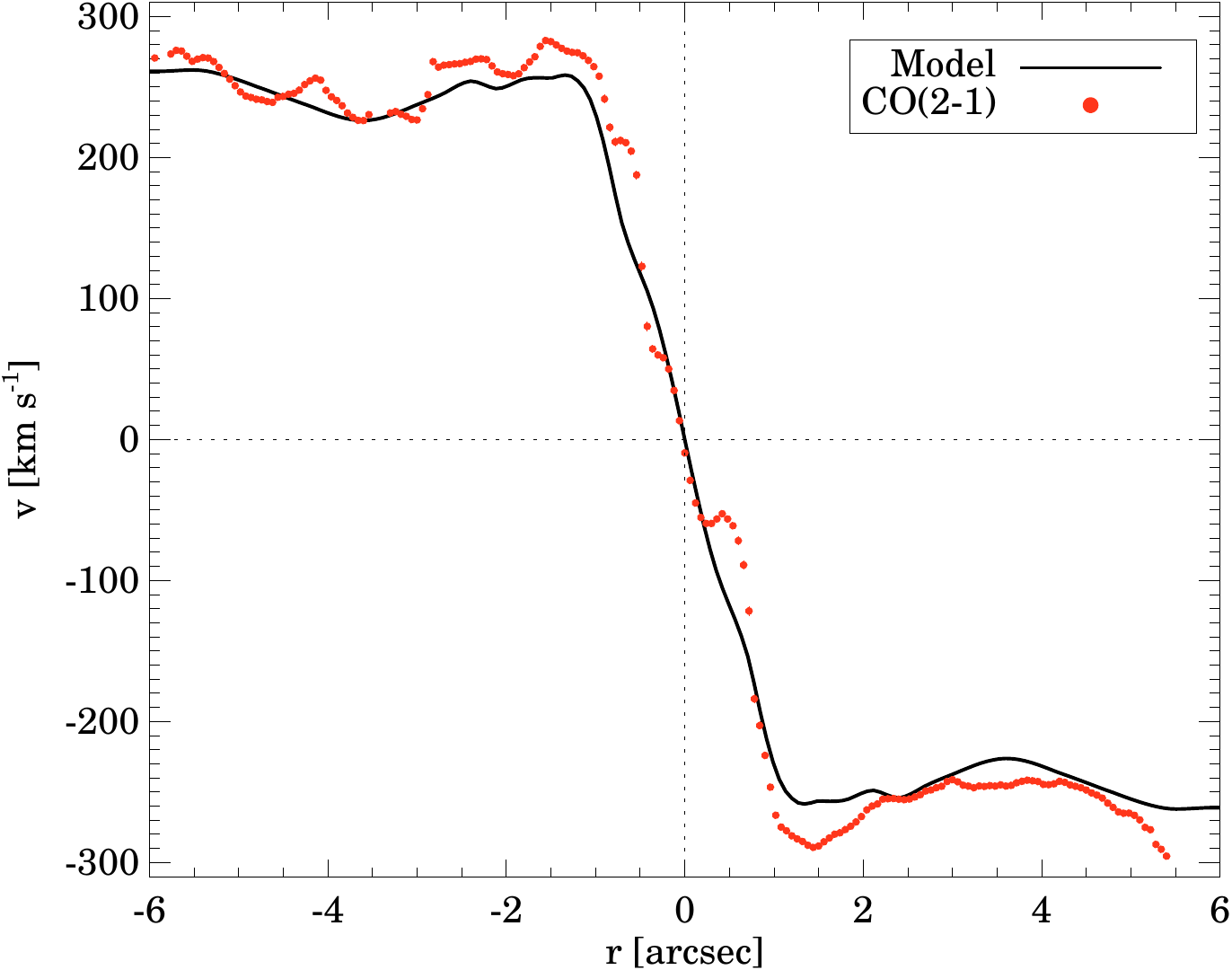}
\caption{\small Rotation curve corrected for inclination ($i$=43\degree) along the major axis (PA$=$133\degree) for the cold molecular gas traced by the CO(2--1) transition (red circles).
The error bars of the CO(2--1) points are 1--2\,km\,s$^{-1}$. The solid black line is the best fit model to the CO(2--1) data (see Section \ref{s:vfield}). \label{fig_rotcurve}}
\end{figure}

For the fit, all the parameters were left free to vary, except the inclination that was fixed at the value derived from the elliptical isophote fitting (43\degree; see Section \ref{s:morphology}). The resulting model (middle panel of Figure \ref{fig_velmodel} and Figure \ref{fig_rotcurve}) reproduces well the overall structure of the velocity field. The residuals standard deviation is 14\,km\,s$^{-1}$ (right panel of Figure \ref{fig_velmodel}). The largest deviations ($\sim\pm$70\,km\,s$^{-1}$) occur at the tips of the secondary bar. If we assume that the spiral arms are trailing, from the velocity field, we can conclude that the NE tip of the bar is the most distant from us. 
The negative (positive) velocity of the residuals at the NE (SW) tip of the bar indicate the presence of non-circular motions. Similar structures found in low-$z$ Seyfert galaxies have been interpreted as gas inflows (\citealt{Diniz2015, SchnorrMuller2016}; see also \citealt{vandeVen2010}).
However, these non-circular motions can also be associated with the elliptical orbits followed by the gas in the bar \citep{Wada1994}. 
Therefore, these radial velocities can only be considered as an upper limit, and the contribution from elliptical orbits to the observed velocity amplitude must be evaluated before estimating the gas inflow rate. 

From this fit, we also obtained the PA of the major kinematic axis (133$\pm$2\degree) and the systemic velocity (3080$\pm$4\,km\,s$^{-1}$ in the LSRK frame). {This is in agreement with the CO(1--0) velocity reported by \citet{Mirabel1990}. 

Figure \ref{fig_rotcurve} shows the rotation curve along the major kinematic axis using the first moment map (Figure \ref{fig_isovel}) and the uncertainty associated with it. From this figure, we obtain that the amplitude of the rotation curve corrected for inclination is $\sim$270\,km\,s$^{-1}$.}

\section{Resolved molecular outflow}\label{s:outflow}

\savebox{\boxmap}{\includegraphics[width=0.42\textwidth]{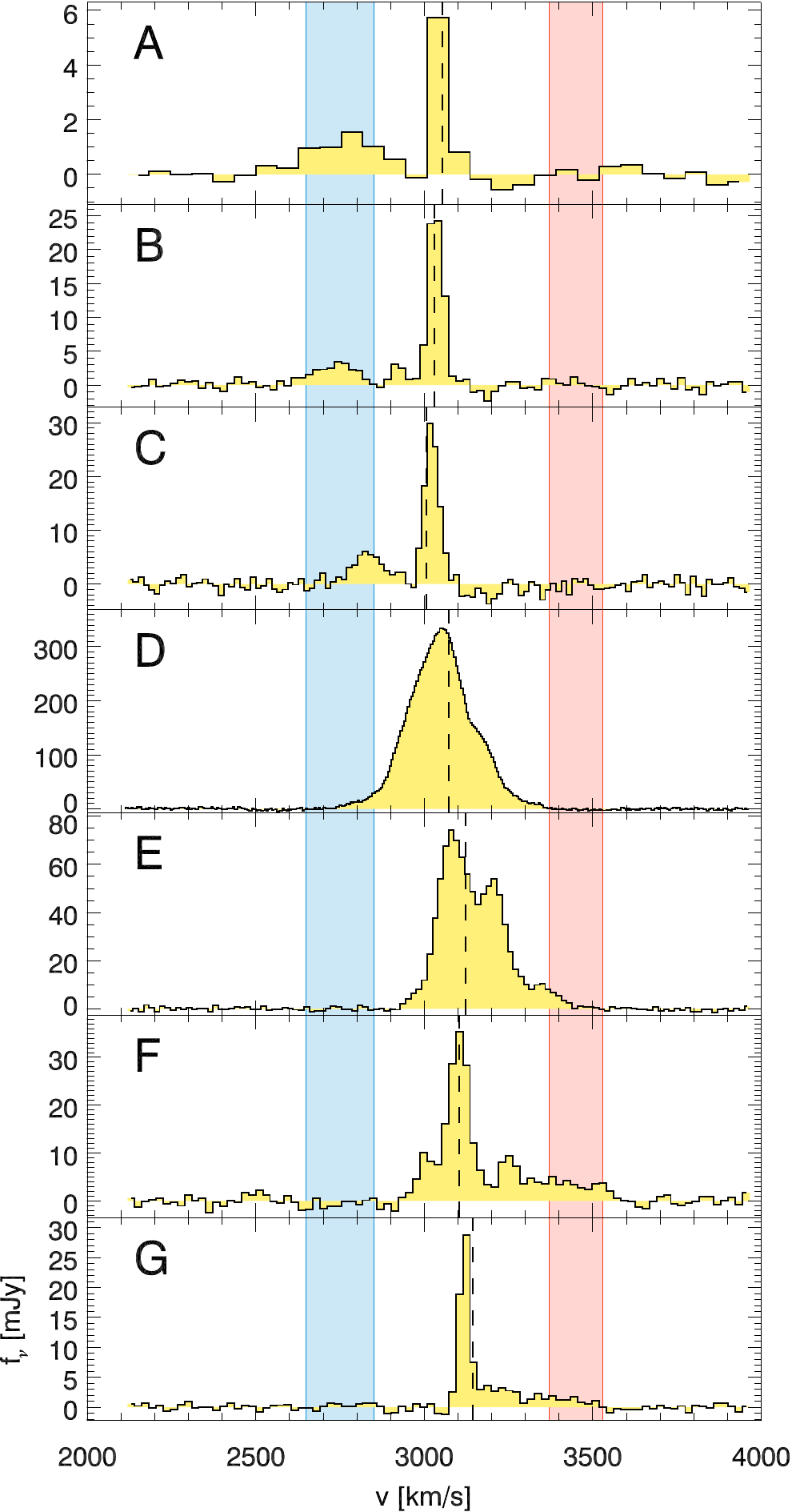}}
\begin{figure*}
\centering
\hfill
\vbox to \ht\boxmap{
\vfill
\hbox to 0.4\textwidth{
\includegraphics[width=0.4\textwidth]{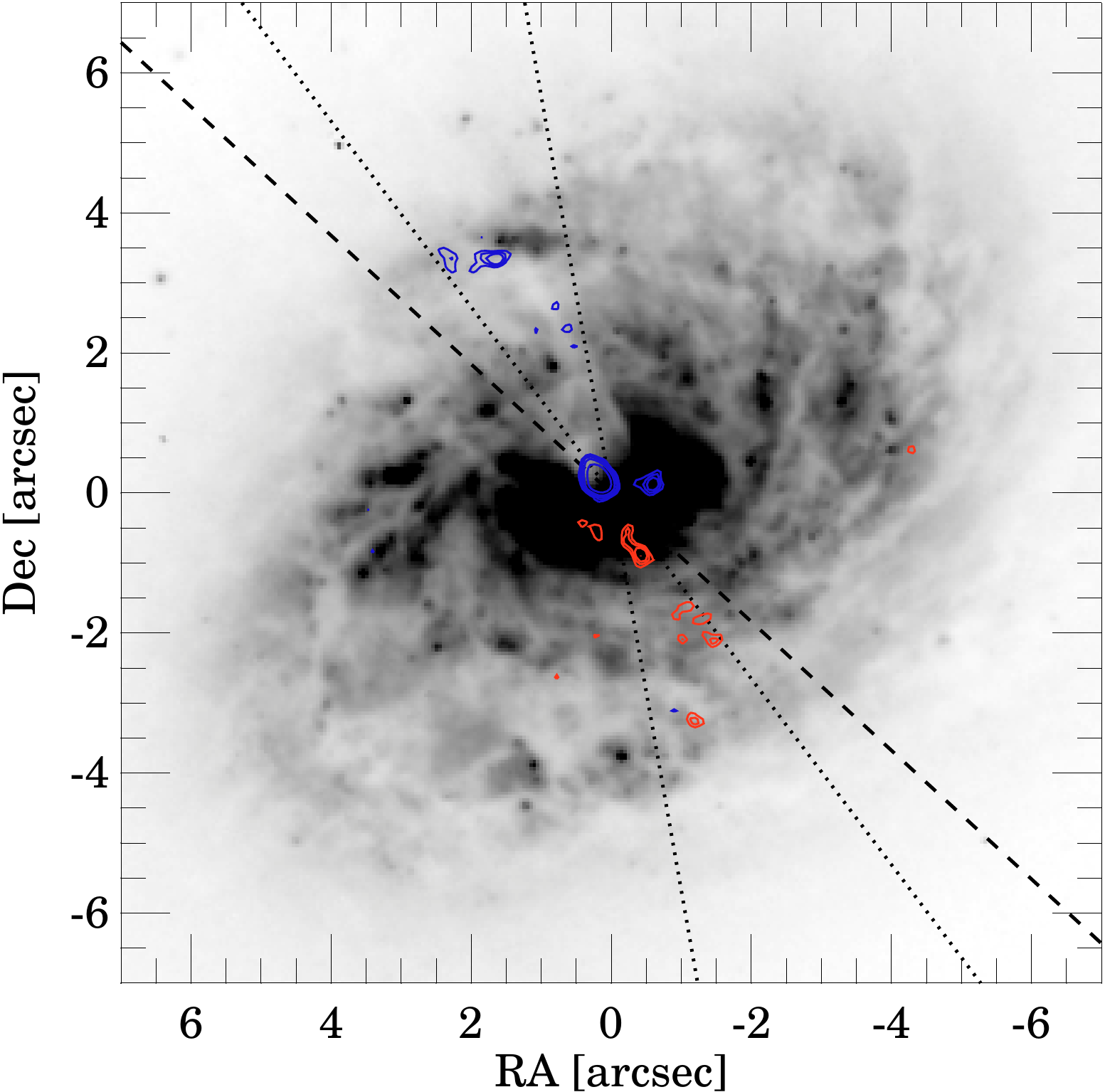}
}
\hbox to 0.4\textwidth{
\includegraphics[width=0.4\textwidth]{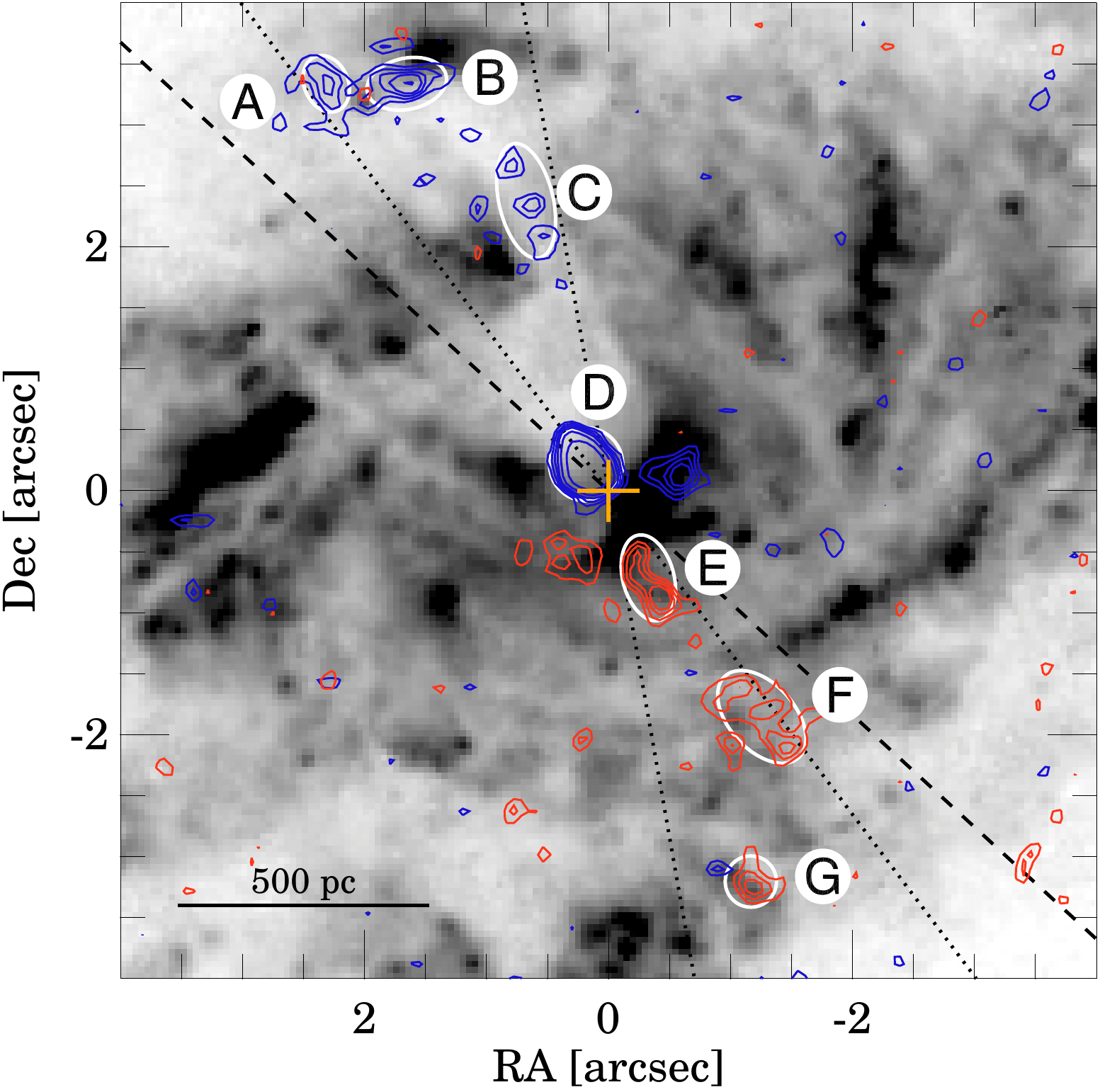}
}
\vfill
}
\hfill
\usebox{\boxmap}
\hfill
\caption{\small The left panels show the CO(2--1) integrated emission between 2650 and 2850\,km\,s$^{-1}$ in blue contours and between 3370 and 3530\,km\,s$^{-1}$ in red contours. The contours correspond to 3, 4, 5, 6, 8, and 12$\sigma$ levels in their respective bands (the 3$\sigma$ contour level is not plotted in the top left panel for clarity). The background of the top left panel is the \textit{HST}\slash ACS F814W image which shows the internal spiral arms. The background of the bottom left panel is the \textit{HST}\slash ACS FR656N image to maximize the contrast of the dust lanes. The dashed line indicates the minor kinematic axis and the dotted lines indicate the approximate opening angle of the outflowing gas. The orange cross marks the position of the 233\,GHz continuum peak.
The right panels show the CO(2--1) spectra of the regions marked in the bottom left panel. The blue and red shaded areas indicate the velocity ranges used to construct the blue and red contours, respectively. The vertical dashed line corresponds to the velocity derived in Section \ref{s:vfield} assuming a rotating disk model. {To increase the signal-to-noise ratio, we binned the spectra in channels of 60, 20, 15, 5, 15, 20, and 20\,km\,s$^{-1}$ from top to bottom.}
\label{fig_outflow}}
\end{figure*}

\begin{table*}[ht]
\caption{Properties of the outflow regions}
\label{tbl_outflow}
\centering
\begin{small}
\begin{tabular}{lcccccccccccc}
\hline \hline
\\
Region & $d\slash \sin i$\tablefootmark{a}  & $v\slash \cos i$\tablefootmark{b} & $\sigma$\tablefootmark{c} & $f$(CO(2--1))\tablefootmark{d} & $\log M_{\rm mol}$\tablefootmark{e} 
& $\log E_{\rm out}$\tablefootmark{f} & $t_{\rm dyn}$\tablefootmark{g} & FWHM\tablefootmark{h} \\
& (pc) & (km\,s$^{-1}$) & (km\,s$^{-1}$) & (mJy\,km\,s$^{-1}$) & ($M_{\rm \odot}$)  & (erg) & (Myr) & (pc) \\
\hline
A &   1320 & --460$\pm$23 & 70$\pm$10 & 250$\pm$10 & 5.5 & 53.8 & 2.8$\pm$0.2 & 70$\pm$20\\
B &   1210 & --510$\pm$16 & 50$\pm$7 & 374$\pm$15 & 5.7 & 54.1 & 2.3$\pm$0.1 & 80$\pm$20\\
C &   800  & --367$\pm$9 & 33$\pm$4 & 474$\pm$13 & 5.8 & 53.9 & 2.1$\pm$0.1 & 70--110\\
D &   90   & [--530,--320]\tablefootmark{*} & \nodata & 1940$\pm$30 & 6.4 & 54.6 & 0.2$\pm$0.1 & 90$\pm$10\\
E & --250  & 370$\pm$11 & 42$\pm$5 & 1044$\pm$15 & 6.1 & 54.2 & 0.7$\pm$0.1 & 150$\pm$30\\
F & --730  & [330, 750]\tablefootmark{*} & \nodata & 920$\pm$40 & 6.1 & 54.5 & 1.4$\pm$0.3 & $\sim$100\\
G & --1100 & [390, 690]\tablefootmark{*} & \nodata & 210$\pm$30 & 5.5 & 53.7 & 2.4$\pm$0.4 & 60$\pm$20\\
\hline
\end{tabular}
\end{small}
\tablefoot{We assume a galaxy inclination of 43\degree\ (see Section \ref{s:morphology}).
\tablefoottext{a}{Deprojected distance.}
\tablefoottext{b}{Deprojected CO(2--1) velocity with respect to the nuclear velocity. When possible, we fitted with Gaussians the line profile of the outflow and the disk emission. Otherwise, we integrated the flux in the given velocity range.}
\tablefoottext{c}{Velocity dispersion derived from the Gaussian fit to the outflow component.}
\tablefoottext{d}{CO(2--1) integrated flux of the outflow component.}
\tablefoottext{e}{Logarithm of the outflowing molecular mass assuming an ULIRG-like conversion factor ($\alpha_{\rm CO}=0.78$ and $r_{21}=0.9$). {If we use the standard conversion factor, the molecular masses will be $\sim$5 times (0.7\,dex) higher.}}
\tablefoottext{f}{Kinetic energy calculated as $1\slash 2\times M_{\rm mol}\times(v\slash \cos i)^2$. For the regions without a Gaussian fit, we used the central velocity of the range. {The uncertainty of the kinematic energy is dominated by the factor of $\sim5$ uncertainty of the molecular mass estimates.}}
\tablefoottext{g}{Dynamical time required for the gas to reach its current position derived using $t_{\rm dyn}=d \times \cot{i} \times v^{-1}$.}
\tablefoottext{h}{FWHM of the regions measured using a 2D Gaussian fit on the outflow emission maps (Figure \ref{fig_outflow}). Region C contains several weak clumps with sizes between 70 and 110\,pc. It was not possible to fit the size of region F, so we indicate the size of the aperture used to extract the spectrum.}
\tablefoottext{*}{The central velocity of the intervals for regions D, F, and G are --425, 540, and 540\,km\,s$^{-1}$, respectively.}
}
\end{table*}

In the ALMA data cube, we identified some regions with secondary CO(2--1) emission separated by $>$200\,km\,s$^{-1}$ from the local modeled velocity in Section \ref{s:vfield}. This deviation is considerably higher than the largest residual ($\sim$70\,km\,s$^{-1}$; see Figure \ref{fig_velmodel}) found for the rotating disk model.
In Figure \ref{fig_outflow}, we show the location of these seven regions and their spectra. The blue- and red-shifted regions are located to the NE and SW of the nucleus, respectively, and approximately symmetric with respect to the nucleus with a PA$\sim$23\degree\ and an opening angle $\sim$30\degree. {Both the spatial distribution approximately along the minor kinematic axis and the high velocities of the molecular gas in these regions are compatible with a molecular outflow almost perpendicular to the rotating disk and with origin at the nucleus of this galaxy.}
Alternatively, these blue- and red-shifted regions could be due to coplanar non-circular motions. Since they are approximate aligned with the minor kinematic axis, these motions would be mainly radial motions. However, the high velocity of this gas (comparable to the rotation curve half amplitude) is much higher than that expected for density-wave-driven gas flows \citep{GarciaBurillo2015}.

{In addition}, there are two regions which also show high blue- and red-shifted velocities close to the nucleus and aligned {approximately perpendicular (PA$\sim$125\degree)} to the outflow. 
We excluded them from the analysis because the rotation and high velocity dispersion of the gas in the inner 400\,pc (see line profiles in Figure \ref{fig_outflow}) can explain their location approximately along the major kinematic axis.

In the optical \textit{HST} images, there is a prominent dust lane that originates at the nucleus in a radial direction close to the positions of the blueshifted regions A, B, C, and D. This is also visible in the F814W\slash F160W ratio (Figure \ref{fig_maps} and also in the near-IR color map presented in \citealt{AAH06s}) where the most obscured region (in red in that ratio map) coincides approximately with region D. These observables {also support the hypothesis of} a molecular dusty outflow originating in the nucleus which produces a screen obscuration over the galaxy disk in our line of sight. As expected in this scenario, we do not find a clear correlation between the optical obscuration and the redshifted part of the outflow (which would be behind the disk in our line of sight).

Figure \ref{fig_outflow} shows that the dust lane associated with the outflow is more or less uniform while the CO(2--1) emission is concentrated in few clumps along the lane.
With the ALMA configurations used to obtain these data, we should be able to recover structures with angular scales up to $\sim$10\arcsec. Thus, it is not likely that diffuse emission from this $\sim$3\arcsec\ region is completely filtered out. On the other hand, we detect weak (3--4$\sigma$) blueshifted CO(2--1) emission between regions C and D, but closer in velocity to the rotation disk model ($\Delta v\sim120$\,km\,s$^{-1}$) and not included in the velocity range considered in Figure \ref{fig_outflow}. This suggests that molecular gas associated with the outflow exists between the selected regions, but with lower velocities and\slash or lower surface brightness and representing just a small fraction of the total CO(2--1) emission of the outflow.
Therefore, the observed clumpy structure of the molecular gas in the outflow agrees with the predictions of numerical simulations where these clumps of cold molecular gas dominates the outflowing molecular gas mass \citep{Zubovas2014, Nayakshin2012}.

\subsection{Physical properties of the clumps}

The PA of the outflow, between 50\degree\ and 80\degree, is relatively close to the minor kinematic axis, (43$\pm$2)\degree, and the opening angle is low (30\degree). Therefore, for simplicity, we assume that the outflow is perpendicular to the galaxy disk when we calculate the deprojected distances and velocities of the outflow regions. 
The CO-to-H$_2$ conversion factor for outflows is poorly known. Therefore, we used the ULIRG-like factor {($\alpha_{\rm CO}\sim0.78$\,\Msun\,(K\,km\,s$^{-1}$\,pc$^{-2}$)$^{-1}$),} which is $\sim$5 times lower than the Galactic conversion factor and {also the ULIRG-like ratio $r_{21}=0.9$ \citep{Bolatto2013}.}
{The molecular gas masses were calculated using Equations 3 and 4 of \citet{Solomon2005} which account for the He mass. The uncertainty in the mass determination is dominated by the factor of $\sim$5 systematic uncertainty in the CO-to-H$_2$ conversion factor. However, the adopted} approach probably provides robust lower limits to the molecular gas mass of the outflow.

\begin{figure}
\centering
\includegraphics[width=0.4\textwidth]{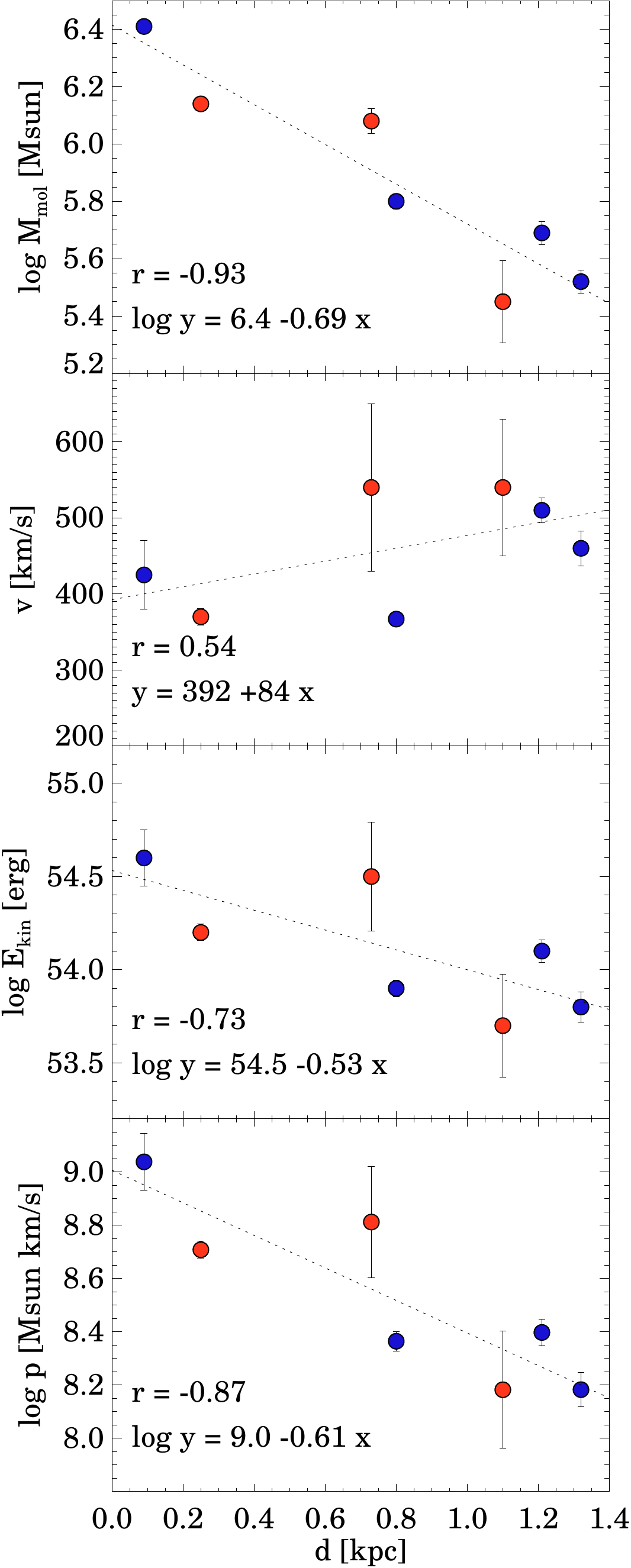}
\caption{Logarithm of the molecular mass (first), velocity (second), kinetic energy (third), and momentum (fourth) of the outflowing clumps as a function of the deprojected distance from the nucleus. Blue and red points are blueshifted and redshifted clumps, respectively. The dotted line is the best fit. The parameters of the best fit and the Pearson correlation coefficients are indicated in each panel. {The error bars correspond to the statistical uncertainties.}
\label{fig_outflow_prop}}
\end{figure}

The properties of the regions are summarized in Table \ref{tbl_outflow}. The total size of the outflow is $\sim$2.5\,kpc and the velocities of the molecular gas with respect to the systemic velocity are $\sim$370--540\,km\,s$^{-1}$ with no clear dependence on the distance from the nucleus (Figure \ref{fig_outflow_prop}).
This implies that the dynamical times required for the gas to reach its current locations are 0.2--3\,Myr.

The sizes of the selected clumps are 60--150\,pc and their masses are in the range 10$^{5.5}$--10$^{6.4}$\,\Msun. Being more massive those clumps closer to the nucleus (top panel of Figure \ref{fig_outflow_prop}). Likewise, the kinetic energy of the clumps, calculated as 1\slash 2 $M_{\rm mol} (v \times \sec i)^2$, and the momentum ($p=M_{\rm mol} v\times \sec i$)
also decrease with increasing distances (bottom panels of Figure \ref{fig_outflow_prop}). Although, these correlations are slightly worse than the mass vs distance relation.
We ignored the kinetic energy associated with turbulence since for those clumps with a well defined $\sigma$ it represents less than 2\%\ of the total kinetic energy.

\subsection{Hot molecular, ionized, and neutral atomic gas}\label{ss:hot_ion_neutral}

Both observations (e.g., \citealt{Dasyra2014, Emonts2014}) and numerical simulations (e.g., \citealt{King2015}) indicate that outflows have a multiphase structure. So far, we have focused our analysis on the new ALMA CO(2--1) data, but we have observed ESO~320-G030 using near-IR and optical integral field spectroscopy \citep{Piqueras2012, Cazzoli2014}. Therefore, it is interesting to use these data to analyze other phases of the outflowing gas.

We used the near-IR $K$-band SINFONI to study the hot molecular phase traced by the H$_2$ 1--0 S(1) 2.12\micron\ transition. $K$-band presents numerous stellar continuum features that can be confused with broad wings of the emission lines. For this reason, we subtracted the stellar continuum spaxel by spaxel using the penalized pixel-fitting method \citep{Cappellari2004} with the NASA Infrared Telescope Facility (IRTF) spectral library \citep{Cushing2005,Rayner2009}.

\begin{figure}
\centering
\includegraphics[width=0.43\textwidth]{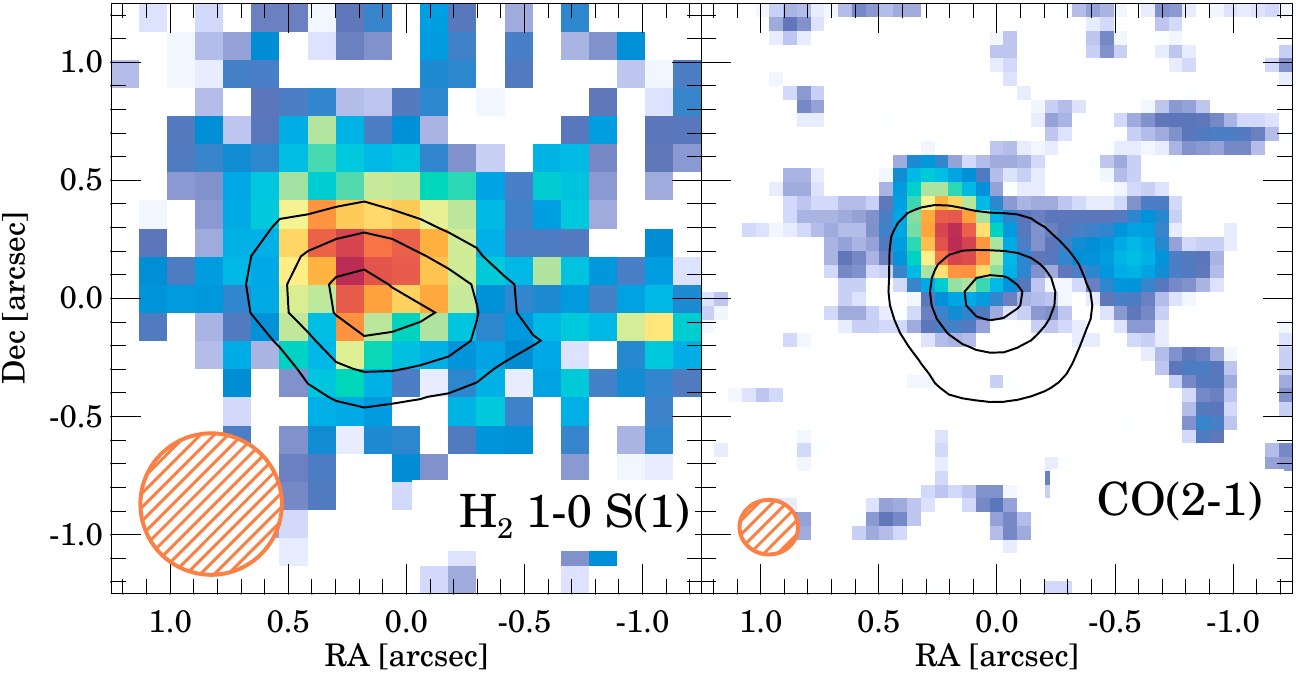}
\includegraphics[width=0.43\textwidth]{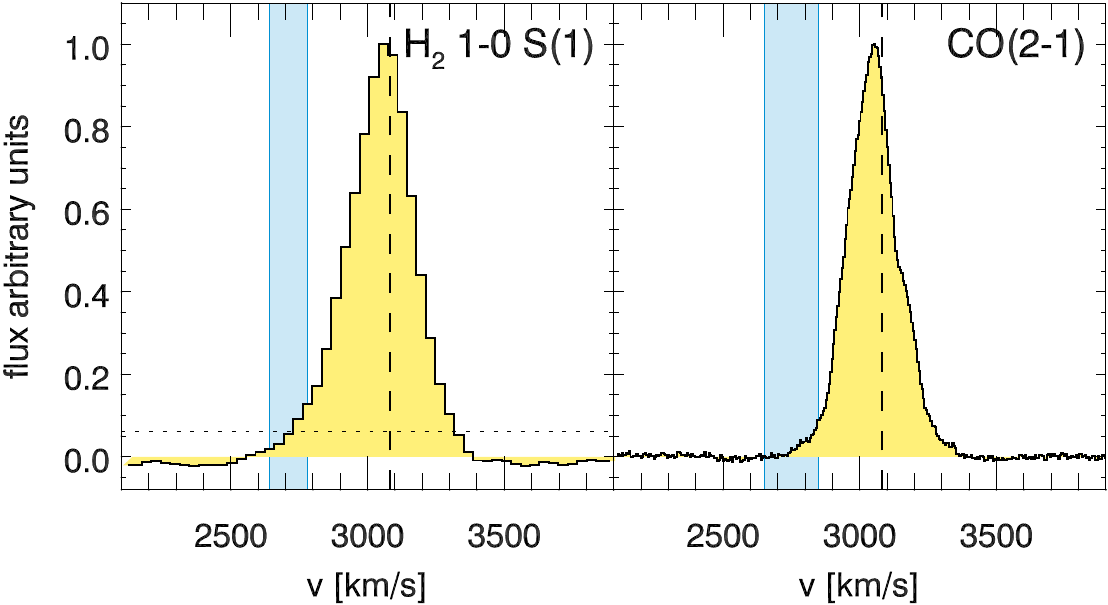}
\caption{Comparison between the hot (left) and cold (right) molecular gas phases of the outflow region D. The hot molecular gas is traced by the H$_2$ 1--0 S(1) 2.12\micron\ transition observed with SINFONI, while the cold molecular gas is traced by CO(2--1). The background images are the integrated emission in the channels shaded in the bottom panels. The contours represent the total emission of the corresponding transition at peak$\times$0.9, $\times$0.6, and $\times$0.3 levels. The orange hatched circles represent the FWHM of the beam of each image (0\farcs6 and $\sim$0\farcs25, respectively). The bottom panels are the spectra extracted from the regions located to the NE of the nucleus in the top panels.
\label{fig_outflow_sinfoni}}
\end{figure}

In the channel maps, we found H$_2$ 1--0 S(1) emission that can be associated with region D. {The rest of near-IR H$_2$ transitions are not detected in this region, probably because they are too weak to be detected with these SINFONI observations.}.
The angular resolution of the SINFONI data is two times worse than the ALMA data and the spectral resolution is $\sim$80\,km\,s$^{-1}$. However, both the velocity range and the location to the NE of the nucleus of the hot H$_2$ emitting region are similar to those of the CO(2--1) region (see Figure \ref{fig_outflow_sinfoni}). Therefore, we consider that they are likely produced in the same physical region.

We measure an H$_2$ 1--0 S(1) flux in this region of (2.5$\pm$0.4)$\times$10$^{-17}$\,erg\,cm$^{-2}$\,s$^{-1}$.
Similar to \citet{Emonts2014}, we assume that the hot molecular gas is thermalized to derive the hot molecular gas mass.
The extinction maps presented by \citet{Piqueras2013} indicate that the nuclear extinction is high. Here, we assume that $A_{\rm k}=4.6$\,mag in the nucleus (Section \ref{ss:nucsfr}). With these numbers, we derive a hot molecular gas mass of 190\,\Msun. 
However, it is uncertain whether the extinction affecting the hot molecular gas is the same affecting the ionized gas traced by {the nuclear} Br$\gamma$. If the extinction is lower, the hot molecular gas mass can be up to tens of times lower.

The ratio of hot-to-cold molecular gas is 7$\times$10$^{-5}$, which is similar to the ratio observed in the AGN powered outflow detected in the LIRG NGC~3256 \citep{Emonts2014}, and is higher than most of the ratios measured in the integrated spectra of local objects including Seyfert galaxies and starbursts (10$^{-8}$--10$^{-5}$; \citealt{Dale2005}).
The enhanced hot-to-cold ratio in NGC~3256 is associated with H$_2$ being heated through shocks or X-rays. In ESO~320-G030, since there is no AGN (see Section \ref{s:nature}), the observed hot-to-cold ratio could be due to shocks in the outflowing gas.

We do not detect near-IR H$_2$ emission in the other clumps of the outflow, but, assuming that the hot-to-cold ratio is also $\sim$1$\times$10$^{-4}$ in those clumps, their hot H$_2$ emission would be below the sensitivity of these VLT\slash SINFONI observations.

To study the ionized component of the outflow, we analyzed the Br$\gamma$ emission line in the SINFONI IFS data {by inspecting the individual channel emission maps}. However, we did not find any hint of emission at the locations and velocities of the molecular outflow. 

Neutral atomic outflowing gas was detected in this object by \citet{Cazzoli2014} analyzing the NaD kinematics in the VIMOS data. A blueshifted NaD absorption component was found to the NE of the nucleus, coincident with the blueshifted molecular gas detected in the ALMA data. Although the morphologies of the neutral atomic and molecular outflowing gas are similar, the velocity range of the neutral atomic gas (30--150\,km\,s$^{-1}$) is lower than that of the molecular gas. 
{However, it is also likely that outflowing gas has a wide distribution of velocities, from few tens of km\,s$^{-1}$ to $>1000$km\,s$^{-1}$ as predicted by numerical simulations (e.g., \citealt{Hopkins2012})}. The derived neutral atomic gas mass of the outflow is 10$^{8.5\pm0.9}$\,\Msun.

\section{Nature of the central source}\label{s:nature}

\begin{figure}
\centering
\includegraphics[width=0.43\textwidth]{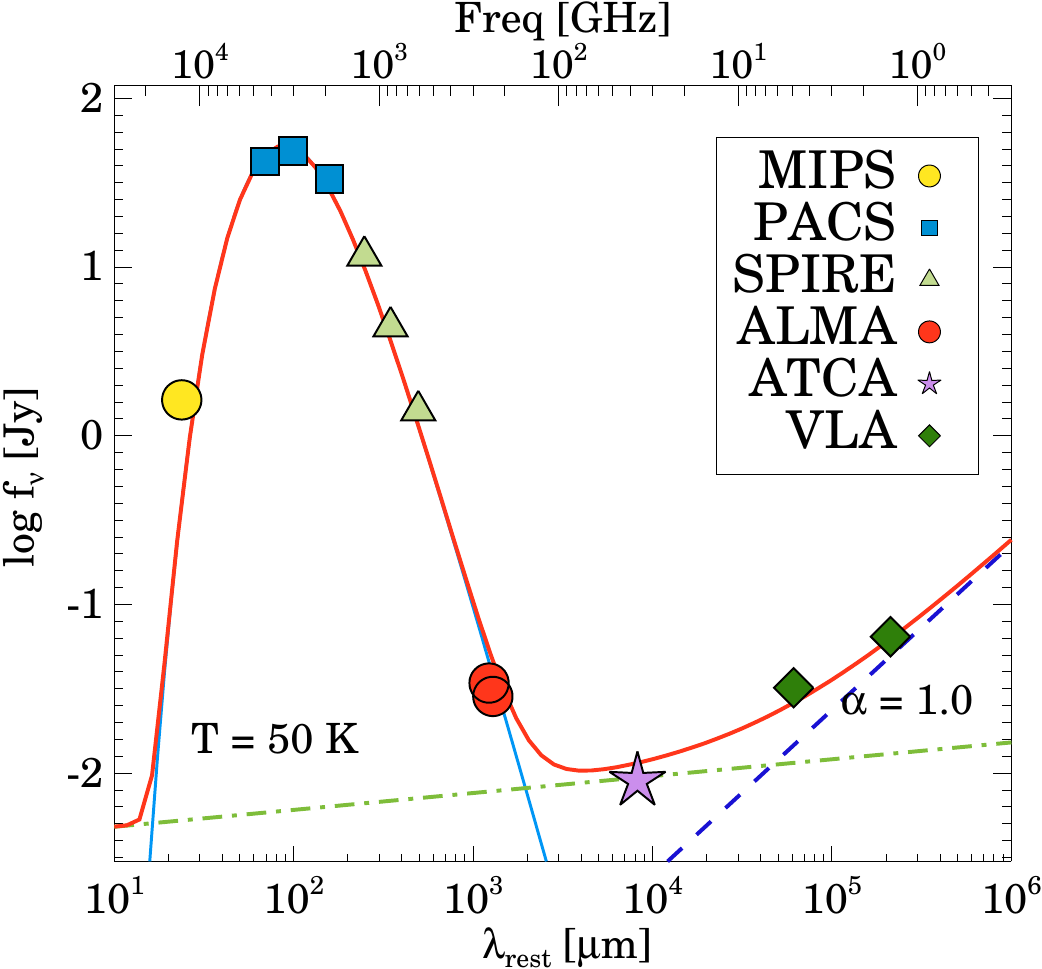}
\caption{Far-IR and radio SED of ESO~320-G030. Tables \ref{tbl_photo_ir} and \ref{tbl_photo_radio} list the wavelength (or frequency), flux, and instrument of the plotted measurements. The cyan line is the best gray body fit to the far-IR photometry (Section \ref{ss:far_ir}), the dashed dark blue line the best power-law fit to the radio ($\nu\leq36$\,GHz) emission after subtracting the thermal radio continuum (dot-dash green line) to the observed data (Section \ref{ss:radio}). The solid red line is the sum of the three components (gray body, thermal radio continuum, and synchrotron). The error bars are smaller than the symbols.
\label{fig_sed}}
\end{figure}

The nuclear region of this object is highly obscured ($A_{\rm k}>$2.3\,mag derived from the 9.7\micron\ silicate absorption; \citealt{Pereira2010}) and even in the near-IR it is not clear where the nucleus is located (see Figure \ref{fig_maps} and also \citealt{AAH06s}). To identify the nucleus, we used the ALMA 230 GHz continuum image where most of the emission is produced by a barely resolved source ({observed size} 0\farcs34$\times$0\farcs30) whose location is compatible with the kinematic center derived from the CO(2--1) velocity field (Section \ref{s:vfield}). This continuum source is also coincident with the peak of the cold and hot molecular gas emissions (CO(2--1) and H$_2$ 1--0 S(1)).

The nuclear activity of ESO~320-G030 is classified as \Hii\ from optical spectroscopy (Appendix B of \citealt{Pereira2011}). Similarly, its X-ray emission is that expected in a starburst \citep{Pereira2011}.
The mid-IR high-ionization transitions of [\ion{Ne}{v}]15.6\micron\ and [\ion{O}{iv}]25.9\micron, which can be used to identify obscured AGNs, are not detected in this object \citep{Pereira2010c}. Moreover, the spectral decomposition of the mid-IR \textit{Spitzer}\slash IRS spectrum of this object did not reveal the presence of an AGN component \citep{AAH2012a}.
Therefore, there is no evidence for an AGN in this object. 

\subsection{Spectral energy distribution}\label{ss:sed}

To further study the power source of the nucleus of ESO~320-G030, we analyzed its spectral energy distribution (SED). In Tables \ref{tbl_photo_ir} and \ref{tbl_photo_radio}, we list the IR and radio measurement used to construct the SED. In addition, to the radio and IR data already published, we measured the ALMA continuum fluxes at 230 and 242\,GHz ($\sim$1.3\,mm), as well as the far-IR fluxes from \Herschel\slash PACS and SPIRE images {(see Section \ref{s:data})}. In all these images, the emission of ESO~320-G030 is barely resolved at their corresponding angular resolutions (from 6 to 35\arcsec; see Table \ref{tbl_photo_ir}). This indicates that the IR emission is dominated by the nucleus and not by the circumnuclear SF regions (see Figure \ref{fig_maps}), at least for $\lambda<100$\micron\ where the angular resolutions are $<7\arcsec$.
This galaxy was recently observed with the Australia Telescope Compact Array (ATCA) at 36\,GHz (B. Emonts 2016, private communication) and we also include this measurement in our SED.

\subsubsection{Radio emission}\label{ss:radio}

First, we fitted with a power-law model ($S_{\nu}\sim \nu^{-\alpha}$) the observed radio emission ($\nu\leq36$\,GHz). The spectral index, $\alpha$, is 0.7$\pm$0.1, which is the value expected for pure synchrotron emission. It is also similar to the values of $\alpha$ found in normal galaxies \citep{Condon1992}. Therefore, the radio emission of ESO\,320-G030 does not suggest the presence of an AGN. Using only the VLA fluxes, \citet{Baan2006} estimated a flatter spectral index of $\sim$0.6, although they also consider this object as a starburst. 

To avoid any bias, in the previous paragraph we ignored the thermal radio continuum that would be produced by the nuclear SF. Since the presence of an AGN in this object seems unlikely, we can estimate the nuclear SFR (15\,\Msun\,yr$^{-1}$; see Section \ref{ss:nucsfr}) and derive the associated thermal radio continuum using Equation 3 of \citet{Condon1992}. First, we derive the equivalent H$\alpha$ flux for the nuclear SFR using the \citet{Kennicutt2012} calibration. From this flux, $F$(H$\alpha$) = 1.1$\times10^{-11}$\,erg\,s$^{-1}$\,cm$^{-2}$ and assuming $T_{\rm e}=10000$\,K, we obtain that $S_{\rm thermal}=13.5 \left( \nu\slash {\rm GHz}\right)^{-0.1} {\rm mJy}$.
This thermal continuum coincides with the 36\,GHz observed flux, and it represents $\sim$40\% and $\sim$25\%\ of the 4.8 and 1.4\,GHz VLA fluxes, respectively. Therefore, after subtracting the thermal continuum, the spectral index is even stepper, $\alpha=1.0\pm 0.1$ (Figure \ref{fig_sed}).

If we extrapolate this model at the ALMA continuum frequencies, we obtain that the thermal radio continuum contribution is 8\,mJy, that is, $\sim$30\%\ of the observed ALMA fluxes at 230 and 242\,GHz.

\subsubsection{Far-IR continuum}\label{ss:far_ir}

Following \citet{Kovacs2010}, we fitted the far-IR emission between 70 and 500\micron\ using a single temperature gray body:
\begin{equation}
S(\nu_{\rm obs}, T_{\rm obs}) = \Omega\left( 1 - e^{-\tau}\right) B_{\nu_{\rm obs}}\left(T_{\rm obs}\right),
\end{equation}
where $S$ is the observed flux density, $\Omega$ the physical solid angle, $\tau$ the optical depth, and $B_{\nu}$ the Planck's blackbody law for a temperature $T$. We assume that the optical depth follows the relation $\tau_\nu = \tau_0 \left( \nu\slash \nu_0 \right)^\beta$, with $\beta=1.8$ \citep{Planck2011}.
The best fit parameters are $T=49\pm2$\,K, $\tau_{350\mu m}=0.30\pm0.04$, and $\Omega=(2.4\pm0.3)\times 10^{-11}$\,sr. This corresponds to an IR luminosity of 10$^{11.12}$\,\Lsun\ and an emitting area with a diameter of 260\,pc (assuming an uniform disk). Using Equation 2 of \citet{Kovacs2010}, we derive a dust mass of 10$^{7.4}$\,\Msun, assuming $\kappa_{350\mu m} = 1.92$\,cm$^2$\,g$^{-1}$ \citep{Li2001}.

The model predicts a flux at 230\,GHz $\sim$30\%\ higher than the measurement and indicates that the emitting area is larger than the area derived from the ALMA continuum map ($\sim$50\,pc), so it is likely that part of the more spatially extended continuum emission is filtered in the ALMA data.

The predicted flux at 24\micron\ is 4 times lower than the observed flux. This indicates that dust with $T>49$\,K exists in the nucleus of this galaxy as expected in a SF object \citep{Rieke2009}. 

To calculate the total IR luminosity, we must include this contribution from warm dust. From the analysis of the SED of local LIRGs, the fraction of the IR luminosity produced by dust with $T<50$\,K is $\sim$80\% \citep{Pereira2015not, daCunha2010}. In addition, the IR emission from polycyclic aromatic hydrocarbons (PAH) is $\sim$10\%\ of the total IR luminosity in LIRGs \citep{Pereira2015not}. Taking into account these two contributions, we derive a total nuclear IR luminosity of 10$^{11.26}$\,\Lsun. This value is in very good agreement with the 8--1000\micron\ IR luminosity derived using the \textit{IRAS} fluxes and our adopted luminosity distance (10$^{11.28}$\,\Lsun; \citealt{SandersRBGS}).

\subsection{Nuclear and circumnuclear star-formation}\label{ss:nucsfr}

The SFR of the circumnuclear regions (1.5\arcsec $<r<$ 5\arcsec) can be directly estimated from the Pa$\alpha$ or Br$\gamma$ maps \citep{AAH06s, Piqueras2016}. In particular, we use the Br$\gamma$ map corrected for extinction using the spatially resolved Br$\gamma$\slash Br$\delta$ ratio \citep{Piqueras2016}. The average $A_{\rm k}$ in the circumnuclear region is 1.4\,mag and the total Br$\gamma$ luminosity excluding the central 3\arcsec\ is {(2.4$\pm$0.1)}$\times$10$^{40}$\,erg\,s$^{-1}$. Using the \citet{Kennicutt2012} SFR calibration for H$\alpha$ {(their Equation 12)} and that H$\alpha$\slash Br$\gamma$ = 104 {for case B conditions, $T=10000$\,K, and $n_{\rm e}=10^3$\,cm$^{-3}$} \citep{Hummer1987}, we derive a circumnuclear SFR of {13$\pm$1}\,\Msun\,yr$^{-1}$. This is similar to the value derived by \citet{RodriguezZaurin2011} using the optical H$\alpha$ emission, 11$\pm$3\,\Msun\,yr$^{-1}$.

The nucleus is almost completely obscured in the Br$\gamma$ map, so we use an indirect method to derive the nuclear SFR. The SFR from the total IR luminosity is 28\,\Msun\,yr$^{-1}$ using the \citet{Kennicutt2012} calibration. Therefore, we estimate that the nuclear SFR is the difference between the total SFR from the IR luminosity and the circumnuclear SFR, that is, {15$\pm$2}\,\Msun\,yr$^{-1}$. We note, that the total IR emission is dominated by the nucleus (see Section \ref{ss:sed}), so it is reasonable that it also dominates the total SFR.

We can now calculate the extinction affecting the nucleus by comparing the observed Br$\gamma$ flux with the expected one from a region with SFR of $\sim$15\,\Msun\,yr$^{-1}$. This gives an extinction of $A_{\rm K} = 4.6$\,mag ($A_{\rm V} = 40$\,mag). Due to this extreme extinction level, the nucleus is almost indistinguishable in the Pa$\alpha$ and Br$\gamma$ maps. 

In this scenario, the cm radio emission is mainly due to nonthermal emission from supernovae remnants {\citep{Condon1992}}. In particular, the 4.85\,GHz ($\sim$6\,cm) emission is related to the SN rate ($\nu_{\rm SN}$) according to
\begin{equation}
\nu_{\rm SN} {\rm (yr^{-1})} = 0.27\times L_{\rm 4.8\,GHz} {\rm (W\,Hz^{-1})} \times 10^{-22}
\end{equation}
assuming that the nonthermal spectral index is 0.8 \citep{Condon1992,Colina1992}. Taking into account that 40\%\ of the nuclear 4.85\,GHz emission is produced by thermal radio emission (Section \ref{ss:radio}), we derive a SN rate of {0.14$\pm$0.02}\,yr$^{-1}$. Another SN rate indicator is the near-IR [\ion{Fe}{ii}]1.64\micron\ emission (e.g., \citealt{Colina1993, AAH2003b}). We measured the nuclear [\ion{Fe}{ii}]1.64\micron\ flux {from the SINFONI data} and corrected it for extinction assuming the $A_{\rm K}$ derived for the ionized gas. The derived SN rate is {1.2$\pm$0.1}\,yr$^{-1}$ using Equation 3 of \citet{AAH2003b}. Although, as for the hot molecular gas, it is uncertain if this value of $A_{\rm K}$ applies to the shocked gas traced by the [\ion{Fe}{ii}].

The SN rate is directly determined by the SFR. For the assumed \citet{Kroupa2001} initial mass function, $\nu_{\rm SN} {\rm (yr^{-1})}=0.012\times$SFR(\Msun\,yr$^{-1}$). Therefore, the expected SN rate is {0.18$\pm$0.02}\,yr$^{-1}$, which is in agreement with the rate derived from the radio emission, but it is 7 lower than that derived from the [\ion{Fe}{ii}] emission. Because of the extreme nuclear obscuration in this object, the latter is highly dependent on the assumed extinction, so this comparison suggests that the nuclear [\ion{Fe}{ii}] emission is less affected by extinction than the ionized gas ($A_{\rm K}=3$\,mag for the [\ion{Fe}{ii}] emission would reconcile the three SN rate estimates).

To determine the star-formation efficiency, we estimate the molecular gas in the nucleus from the CO(2--1) ALMA data. The CO(2--1) emission is more extended than the 232\,GHz continuum (0\farcs74$\times$0\farcs65, 180\,pc$\times$160\,pc). This size is more similar to that derived from the far-IR SED (260\,pc; Section \ref{ss:far_ir}), so it seems likely that the nuclear SF occurs in a region with a diameter between 170 and 260\,pc. Therefore, to measure the nuclear CO(2--1) flux, we use an aperture with $d=1\farcs2$ ($\sim$290\,pc). The resulting flux is 202.6$\pm$0.1\,Jy\,km\,s$^{-1}$, that corresponds to a molecular gas mass of 10$^{9.1}$\,\Msun\ using the Galactic CO-to-H$_2$ conversion factor which is appropriate for SF regions in LIRGs \citep{Pereira2016}. Therefore, the gas-to-dust ratio in the nucleus is $\sim$50--60, which lies at the lower end of the observed ratios in U\slash LIRGs \citep{Wilson2008}.
The depletion time ($t_{\rm dep}=M_{\rm H_2}\slash {\rm SFR}$) of the nucleus is $\sim$80\,Myr. This $t_{\rm dep}$ is much shorter than those measured in normal spiral and Seyfert galaxies ($\sim$1\,Gyr; \citealt{Leroy2013, Casasola2015}), but we found similar $t_{\rm dep}$ values in individual regions of another LIRG (IC~4687) observed at a similar spatial resolution \citep{Pereira2016}. Therefore, the nucleus of ESO~320-G030 is not particularly extreme in terms of SF efficiency.

However, the molecular gas surface density is 10$^{4.4}$\,\Msun\,pc$^{-2}$, which is $\sim$20 times higher than the highest surface densities observed in local objects and also $\sim$10 times higher than in the regions of IC~4687 (see Figure 7 of \citealt{Pereira2016}). Similarly, the SFR surface density is 10$^{2.5}$\,\Msun\,kpc$^{-2}$, which is $\sim$2 orders of magnitude higher than in normal galaxies and 3 times higher than the most extreme value of IC~4687. To find similar values of molecular gas and SFR surface densities, we need to go to merger LIRGs and ULIRGs (e.g., \citealt{Xu2015}).

\begin{table}[ht]
\caption{Infrared photometry of the nucleus}
\label{tbl_photo_ir}
\centering
\begin{tiny}
\begin{tabular}{lcccccccccccc}
\hline \hline
& & & Beam & Nuclear\\
Name & $\lambda_{\rm obs}$ & $f_\nu$ & FWHM & FWHM\tablefootmark{a} & Ref. \\
& $(\micron)$ & (Jy) & (\arcsec) & (\arcsec) & \\
\hline
MIPS & 24  & 1.72$\pm$0.02 & 5\farcs9 & 7\farcs1 &1\\
PACS & 70 & 42.3$\pm$0.2  & 5\farcs6 & 6\farcs4 &2\\
PACS & 100  & 48.6$\pm$0.2 & 6\farcs8 & 7\farcs7 &2\\
PACS & 160 & 33.4$\pm$0.1 & 11\farcs4 & 12 &2\\
SPIRE & 250  & 11.7$\pm$0.1& 17\farcs6 & 18 &2\\
SPIRE & 350  & 4.4$\pm$0.1 & 23\farcs9 & 26 &2\\
SPIRE & 500  & 1.42$\pm$0.02 & 35\farcs2 & 37 &2\\
\hline
\end{tabular}
\end{tiny}
\begin{small}
\tablefoot{{Nuclear fluxes of ESO~320-G030 from the \Spitzer\slash MIPS, \Herschel\slash PACS, and \Herschel\slash SPIRE observations. In all these images, the integrated emission of ESO~320-G030 is dominated by a barely resolved nuclear source.}
\tablefoottext{a}{Measured nuclear FWHM using a 2D Gaussian fit.}}
\tablebib{(1) \citealt{Pereira2011}; (2) This work.}
\end{small}
\end{table}

\begin{table}[ht]
\caption{Radio photometry of the nucleus}
\label{tbl_photo_radio}
\centering
\begin{tiny}
\begin{tabular}{lcccccccccccc}
\hline \hline
\\
Name & $\nu_{\rm obs}$ & $f_\nu$ & Beam & Ref. \\
& (GHz) & (mJy) & (\arcsec)$\times$(\arcsec) & \\
\hline
ALMA & 230 & 28.5$\pm$0.1 & 0.25$\times$0.23 & 1\\
ALMA & 242 & 34.2$\pm$0.2 & 0.25$\times$0.23 & 1\\
ATCA & 36 &  8.8$\pm$0.1 & 3.2$\times$1.4 & 2\\
VLA & 4.85 & 32.1$\pm$0.4 & 0.36$\times$0.27 & 3 \\
VLA & 1.4  & 64.4$\pm$0.8 & 2.2$\times$0.6 & 3 \\
\hline
\end{tabular}
\end{tiny}
\begin{small}
\tablebib{(1) This work; (2) B. Emonts 2016, private communication; (3) \citealt{Baan2006}}
\end{small}
\end{table}

\section{Discussion}\label{s:discussion}

In this section, we discuss several aspects of the massive molecular outflow detected in ESO~320-G030.

\subsection{Physical properties and kinematics}

We measured a molecular gas mass of 10$^{6.8}$\,\Msun\ in the outflow clumps. Using the Galactic CO-to-H$_2$ conversion factor, this mass would be 5 times higher, so this number is a lower limit. Moreover, it is possible  that diffuse emission exists between these clumps, although, according to models, {most of the gas of outflows is in fragmented clouds \citep{Nath2009}.}

On average, the molecular mass outflow rate is $\dot{M}_{\rm out}=$1.7\,\Msun\,yr$^{-1}$ (assuming a 3\,Myr dynamical time). This corresponds to a loading factor ($\dot{M}_{\rm out}\slash {\rm SFR}$) $\sim$0.1 (or 0.5 using the Galactic CO conversion factor).
Loading factors {about 0.2--3} are often observed in starbursts \citep{Salak2016, Cicone2014}, so the observed loading factor is compatible with a star-formation origin as suggested by the nuclear activity. 
Although, this low loading factor indicates that star-formation quenching is not efficient in this object.

The velocity of the outflowing gas is $\sim$370--540\,km\,s$^{-1}$ (see Section \ref{s:outflow}) and
the escape velocity at 1.5\,kpc for this object is 530\,km\,s$^{-1}$ (see \citealt{Cazzoli2016}). Therefore, this suggests that most of this outflowing molecular gas will return to the disk after several Myr. This recycling of material is also compatible with the slowly rotation thick disk of neutral atomic gas found in this object \citep{Cazzoli2014}.

\subsection{Origin and evolution of the clumpy structure}

In Section \ref{s:outflow}, we showed that the properties of the clumps change with the distance from the nucleus. 
Two possibilities can explain the presence of molecular gas in outflows: (1) it forms in overdensities in the outflowing neutral\slash ionized gas (e.g., \citealt{Zubovas2014b}); or (2) the molecular gas is dragged from the molecular phase of the interstellar medium (ISM; e.g., \citealt{Hopkins2012}). In the first case, despite the clumpy structure of the outflow, the outflow rate can be relatively constant with time, while in the second case, the clumps might correspond to peaks of the outflow rate.

From Figure \ref{fig_outflow_prop}, we see that the mass of the clumps decreases with the distance from the nucleus, but the velocity remains approximately constant. If the clumps appear in neutral atomic gas overdensities, this indicates that, further from the outflow origin, these overdensities are less common or that the formation of the molecular gas is less efficient. The volume occupied by the outflow increases proportionally to $d^3$, so both the larger volume of the outflow and the lower gas density at higher distances from the origin would agree with this possibility. On the other hand, if the clumps were originally part of the ISM, they would be losing mass, energy, and momentum as they move. The loss of energy and momentum is expected due to turbulence and the effect of the gravity potential of the galaxy. The mass loss can be due to the clump evaporation in a hot gas outflow environment \citep{Cowie1977}.

\subsection{Power source and energetics}

Outflows are produced in starbursts due to supernova explosions, stellar winds, and radiation pressure
\citep{Hopkins2012}. 
Here, we investigate if SNe alone could power the molecular outflow of ESO~320-G030 by comparing the kinetic energy and momentum provided by SNe and those measured in the outflowing clumps. 

The kinetic energy of the molecular gas clumps is $\sim$10$^{55}$\,erg\,s$^{-1}$, or $\sim$4$\times$10$^{48}$\,erg\,s$^{-1}$\,yr$^{-1}$ averaged over the dynamical time. For the SN rate of the nucleus ($\sim$0.2\,yr$^{-1}$) and assuming an energy of 10$^{51}$\,erg per SN, the kinetic energy associated with this molecular outflow is $\sim$2\% of the total available energy.
Similarly, the total momentum is 3$\times$10$^9$\,\Msun\,km\,s$^{-1}$ and the momentum rate $\sim$10$^3$\,\Msun\,km\,s$^{-1}$\,yr$^{-1}$. The radial momentum per SN is 2.8$\times$10$^4$\,\Msun\,km\,s$^{-1}$ \citep{Walch2015}, so about 20\%\ of the momentum due to SNe is deposited in the outflowing molecular gas. 
Therefore, the observed outflow can be powered by SNe, although stellar winds and radiation pressure might contribute as well.

\section{Conclusions}\label{s:conclusions}

We present high spatial resolution ($\sim$60\,pc) ALMA CO(2--1) observations of the local spiral LIRG ESO~320-G030. We study the morphology and kinematics of the cold molecular gas traced by the CO(2--1) emission and combine these data with ancillary \textit{HST} optical and  near-IR imaging as well as VLT\slash SINFONI near-IR integral field spectroscopy. The main goals of this work are: characterize the resolved massive molecular outflow detected in this object and establish the nature of the extremely obscured nucleus which produces the massive molecular outflow.
The main results are the following:

\begin{enumerate}

\item The global kinematics is well represented by a regular rotating disk, although we find non-circular molecular gas motions related to the secondary bar. This can be the signature of inflowing gas motions, which might explain the extreme concentration of molecular gas in the central 500\,pc of ESO~320-G030 (60\% of the total CO(2--1) emission and 10$^{4.4}$\,\Msun\,pc$^{-2}$).

\item We spatially resolve a high velocity ($\sim$450\,km\,s$^{-1}$) outflow containing 10$^{6.8}$\,\Msun\ of molecular gas (assuming the ULIRG conversion factor) originating in the central $\sim$250\,pc. The size of the outflow is $\sim\pm$1.2\,kpc, which corresponds to a dynamical time of $\sim$2.8\,Myr. The opening angle is $\sim$30\degree.

\item We measure the properties of 7 clumps in the outflow. Their sizes are 60--150\,pc and they have molecular gas masses between 10$^{5.5}$ and 10$^{6.4}$\,\Msun (assuming an ULIRG-like conversion factor). The mass, kinetic energy, and momentum of the clumps decrease with increasing distances while the velocity is approximately constant. 

\item The hot molecular gas component of the outflow, as probed by the near-IR H$_2$ transitions, is detected in the innermost ($\sim$100\,pc) part of the outflow with a hot-to-cold molecular gas ratio of 7$\times$10$^{-5}$. This ratio is similar to that measured by \citet{Emonts2014} in another resolved molecular outflow.

\item We find that the nuclear IR and radio emission of the nucleus ($d\sim250$\,pc) are compatible with highly obscured intense SF ($A_{\rm k}\sim$\,4.6\,mag; SFR\,$\sim$\,15\,\Msun\,yr$^{-1}$). No evidence for the presence of an AGN is found. The outflow rate and loading factor are $\dot{M}_{\rm out}=$2--8\,\Msun\,yr$^{-1}$ and $\sim$0.1--0.5, respectively, and depending on the CO conversion factor assumed. This low loading factor indicates that star-formation quenching due to the molecular outflow is not efficient in this object.

\item We find that SN explosions in the nuclear starburst ($\nu_{\rm SN}\sim$0.2\,yr$^{-1}$) can power the observed molecular outflow. The kinetic energy and radial momentum of the outflow represent $\sim$2\% and 20\%, respectively, of the SNe output. 

\item The velocity of the outflowing clumps is lower than the escape velocity, so it is likely that most of this molecular gas will return to the disk after several Myr. This is compatible with the thick neutral atomic gas disk found in this object.
\end{enumerate}

{The origin of this outflowing molecular gas (either formed in overdensities in the outflowing ionized\slash neutral gas or dragged from the nuclear ISM) is not clear from the available data. New high-spatial resolution observations of the different phases of this outflow will help to establish the formation and evolution of the observed molecular gas.
}

\begin{acknowledgements}
{We thank the anonymous referee for useful comments and suggestions.}
We acknowledge support from the Spanish Plan Nacional de Astronom\'ia y Astrof\'isica through grants AYA2010-21161-C02-01 and AYA2012-32295.
AA-H acknowledges financial support from the Spanish Ministry of Economy and Competitiveness through grant AYA2015-64346-C2-1-P. BE acknowledges funding through the European Union FP7-PEOPLE-2013-IEF grant 624351
This paper makes use of the following ALMA data: ADS/JAO.ALMA\#2013.1.00271.S. ALMA is a partnership of ESO (representing its member states), NSF (USA) and NINS (Japan), together with NRC (Canada) and NSC and ASIAA (Taiwan), in cooperation with the Republic of Chile. The Joint ALMA Observatory is operated by ESO, AUI/NRAO and NAOJ.

\end{acknowledgements}

\def\aj{AJ}%
\def\araa{ARA\&A}%
\def\apj{ApJ}%
\def\apjl{ApJ}%
\def\apjs{ApJS}%
\def\aap{A\&A}%
\def\aapr{A\&A~Rev.}%
\def\aaps{A\&AS}%

\end{document}